\documentclass[letterpaper,prd,showpacs,showkeys,10pt,twoside,byrevtex,nofootinbib,superscriptaddress]{revtex4-1}

\usepackage[latin1]{inputenc}
\usepackage[english]{babel}
\usepackage{amsmath}
\usepackage{graphics}
\usepackage{graphicx}
\usepackage{hyperref}
\usepackage{color}

\newcommand{\beq}{\begin{equation}}
\newcommand{\eeq}{\end{equation}}
\newcommand{\beqq}{\begin{equation*}}
\newcommand{\eeqq}{\end{equation*}}
\newcommand{\barr}{\begin{eqnarray}}
\newcommand{\earr}{\end{eqnarray}}
\newcommand{\bea}{\begin{eqnarray*}}
\newcommand{\eea}{\end{eqnarray*}}

\newcommand{\dphi}{\delta \phi}
\newcommand{\mH}{\mathcal{H}}

\newcommand{\x}{\textbf{x}}
\newcommand{\y}{\textbf{y}}

\newcommand{\bra}{\langle}
\newcommand{\ket}{\rangle}

\newcommand{\half}{\frac{1}{2}}

\begin{document}
\title{ CSL Wave Function Collapse Model as a Mechanism for the Emergence of Cosmological Asymmetries in Inflation}
\author{Pedro \surname{Ca\~{n}ate}}
\email[E-mail:\,]{pedro.canate@nucleares.unam.mx}
\affiliation{Instituto de Ciencias Nucleares, UNAM \\
  M\'exico D.F. 04510, M\'exico}
\author{Philip  \surname{Pearle}}
 \email[E-mail:\,]{ppearle@hamilton.edu}
\affiliation{Department of Physics, Hamilton College \\
Clinton NY 13323, USA}
\author{Daniel \surname{Sudarsky}}
\email[E-mail:\,]{sudarsky@nucleares.unam.mx}
\affiliation{Instituto de Ciencias Nucleares, UNAM \\
  M\'exico D.F. 04510, M\'exico}


\begin{abstract}

 As  previously discussed  in \cite{shortcomings}, the  inflationary  account  for   the  emergence of the seeds of cosmic  structure falls  short of  actually
   explaining the   generation  of  primordial anisotropies and  inhomogeneities. This description  starts  from  a  symmetric  background,   and  invokes  symmetric   dynamics, so it 
 cannot explain asymmetries. To generate asymmetries, 
we present an application of the Continuous Spontaneous Localization  (CSL) model  of wave function collapse\cite{CSL} in the context of inflation.  
This modification of quantum dynamics introduces  a  stochastic  non-unitary  component to the evolution of  the inflaton field perturbations.    
This leads to passage from a  homogeneous and isotropic stage to another, where the quantum uncertainties  in the  initial state of inflation  transmute into
  the  primordial inhomogeneities and anisotropies. We  show, by proper choice of the collapse-generating operator, that it is possible to achieve  compatibility with the  precise observations  of  the 
cosmic microwave background (CMB) radiation. 

\end{abstract}

\maketitle

\section{Introduction}\label{Intro}

The  measurement problem  in  quantum mechanics   remains, 
 almost a century after the theory's  formulation,  the  major obstacle to considering  the  theory  as truly fundamental.
 Despite  heroic  efforts  by    many  insightful  physicists,  the difficulties   involved   have not yielded  and  we are still lacking  any   fully  satisfactory   option. 
 The  basic issue, as  described  for instance  in \cite{measurement},
 is the  fact that the  theory  relies  on two different and incompatible  evolution  processes.  Using  Penrose's characterization\cite{Penrose}, there is the  $U$ (unitary)  process,  where the state  changes  smoothly  according to Schr\"oedinger's   deterministic  differential  equation, and  the $R$ (reduction) process, in which the state of the system  changes  instantaneously, in an indeterministic   fashion. The  $U$ process  is  supposed to control a system's  dynamics  all the time that  the system is  left  alone,  while  the $R$  process is called upon  whenever a measurement  has been  carried  out.
  
 The  problem  is that no one has been able to characterize in general when a physical process should be considered  a {\it measurement}.  This issue has  been  studied and  debated extensibly  in the scientific and  philosophical  literature \cite{More measurement}.  Of course, in laboratory situations,  one clearly knows  when a measurement has been carried out.   Nonetheless, as characterized by J. Bell \cite{Bell}, a {\it for all practical purposes} (FAPP)  approach  
is  not satisfactory  at the foundational level, as it involves treating  the system differently from the measuring device 
or the observer, and this division is one for which the theory offers no specific internal rules. 
 Its  most  conspicuous  
inadequacy occurs in cosmological applications, where  one can not  use any interpretation that relies on  an  observer,  or  on a  measurement  device.\cite{QMCosmolgy},

One approach\cite{Collapse theories} to resolving this problem of standard quantum theory is to modify  it  by incorporating  novel  dynamical features  responsible for  `the collapse of the wave function'.  It may be  characterized  as  the promotion of quantum theory from a theory of measurement to a theory of reality,  in which some of  the  physical properties of a given system take values, regardless of whether they are observed  or not. Such an approach 
\textit{can} be applied to cosmology, and we shall do so by focusing on the  problem of   emergence of  the seeds of   structure in  inflationary  cosmology.
 
	 According to the   standard  account of inflation, in its early stages, a  relatively generic state  of the universe\footnote{ There are  several  conditions   that  are required   for this, but we  shall not  elaborate on  that  here. } is driven  towards  a   homogenous and isotropic  Robertson  Walker  space-time.  This  expands  almost exponentially, driven by the  potential of  the inflaton field, which  acts  as  a large  effective   cosmological constant. The inflaton field fluctuation is a  quantum field\footnote{ There are  scalar,  vector   and  tensor  fields,  but we  will focus  here  on just the scalar field  which is responsible for the   anisotropies  that  have  been observed   in the CMB, so far.} which  is taken to initially be  in the so called  Bunch-Davies  vacuum  state.   \footnote{The fact that,  due to the  kinetic  term in the energy momentum  tensor of the scalar field,  the  space-time can   not exactly    correspond  to the de Sitter   metric, implies that the  state is not   exactly the  Bunch-Davies state.  However, this difference is negligible for our treatment.}	
 This  state is   completely homogeneous and isotropic, and the dynamics preserves these properties. Therefore, it cannot be used to explain the observed inhomogeneous and anisotropic distribution of the primordial energy density in our universe. There are  various  proposals   to address  this issue.  Oneproposal postulates that the state vector can be expanded in some `natural'  basis and, somehow, it is  one of these basis  states which  describes our inhomogeneous and anisotropic  universe. As discussed in  detail in\cite{shortcomings},  none  of them  can be  regarded  as satisfactory.
  
There are analogous instances  where  quantum theory presents us  with  a symmetric quantum state whereas nature exhibits asymmetric behavior.  Perhaps the  best known example is the $ \alpha$ decay of a $J=0$  nucleus.  Although the wave function is rotationally invariant, the alpha particles are seen to move on linear  trajectories.  The  problem was  studied  by N. Mott\cite{Mott}, and was  resolved by heavy usage of the collapse postulate appropriate to a measurement situation\cite{shortcomings}. 
However,  in  the cosmological problem at hand,  even if we wanted to, we  could not  achieve  a similar explanation,  for  we can not  call upon any external  entity  making a measurement.  

The inflaton field fluctuation, according to the  standard  accounts, is supposed  to describe the seeds of the growth of structure
     and lead to the formation of  galaxy clusters,  galaxies,  stars,   etc. The prediction, which  involves the expectation value of the product of two inflation field operators at the end of the inflationary period, is phenomenologically quite successful.  
But, as we  have argued, in order to be   truly  successful   those  accounts  should also   explain  the  actual emergence of   inhomogeneities and  anisotropies
     from the   quantum  uncertainties of a  quantum state that is   completely homogeneous and isotropic.  The issue  is sometimes referred  as  that of the ``classicalization of the fluctuations".
         There  have  been important efforts in this  direction, among which are  works  like\cite{Referee 1} and  \cite{Referee 2}   which  focus  on  the squeezing of the quantum  state of the inflaton fluctuations as  a result of the cosmological expansion as  well  as  others  that focus on  the  role of decoherence \cite{Decohe}, as  well as  works  where  both  aspects  are emphasized
         \cite{kiefer}. As explained in detail in  \cite{shortcomings},  those  approaches are  not fully satisfactory.
  Basically,  there is no way we can invoke  anything like a measurement  postulate  as  part of the explanation of the
  emergence of those   primordial seeds   of structure.  Not  only   there  were no  observers  or anything to play the role of a measuring apparatus  at the time but,  because  we  and our  measuring devices  (and indeed  any conceivable  kind of   astronomers of    alien  civilizations)   owe  their  existence  to the  process  generating those  seeds, they  can  not be  any  part of   their cause!  A  careful reading of,  say  \cite{Referee 1}, uncovers  the important role that measurements  would  have to play in  any such account.

To summarize, the problem is that one cannot explain the emergence of  the observed asymmetries in homogeneity and isotropy when one has a theory with an initial state,  the  Bunch-Davies  vacuum on  the  RW  space-time,   which is $100\%$ homogeneous and isotropic and  a dynamics  that does not break such symmetries.  Nonetheless, we wish to 
 recover the phenomenological success of  
 the  standard  account, where  quantum fluctuations in the vacuum state are the source  of   the first  inhomogeneities  and  anisotropies.\footnote{We  emphasize that the so-called  quantum fluctuations  are  quantum uncertainties  and  should not be  confused  with  thermal  or statistical fluctuations.}
  
 In the case of alpha-decay, dynamical collapse theories, either the GRW\cite{GRW} model or the CSL\cite{CSL} model, satisfactorily resolves the problem.  In CSL, which is usually considered to have superseded GRW, the Schr\"odinger equation is modified by adding to the usual hamiltonian a non-unitary term $iH_{C}$, where $H_{C}$ is a unitary `collapse hamiltonian.'  $H_{C}$ depends upon a randomly fluctuating classical field $w(\bf{x},t)$. The eigenstates of  $H_{C}$ are essentially mass density eigenstates.   This modified 
  Schr\"odinger equation  is supplemented by a second equation, the Probability Rule, which gives the probability that nature chooses a particular $w(\bf{x},t)$. The dynamics is such that, for each $w(\bf{x},t)$ of high probability, the collapse hamiltonian evolves the state vector toward one or another eigenstate of mass density, and  this  occurs  according to the Born probability rule.   
 
 In the description of alpha decay, under the usual hamiltonian,  the state vector describes the alpha particle interacting with the gas atoms in its path. 
 With the added CSL dynamics, as more and more gas atoms get involved,  for each high probability $w(\bf{x},t)$, the collapse hamiltonian more and more rapidly  drives the state vector toward an approximate mass-density eigenstate describing the alpha moving in a straight-line path among its associated atoms.  
 
Returning to  the  standard  inflationary scenario,  since there is  no physical  mechanism  in standard quantum theory that  can account for  the  emergence  of the   inhomogeneities  and  anisotropies, it is natural to consider addition of a physical mechanism 
that can do so.  The mechanism of wave function collapse in the cosmological setting was suggested and modeled in an ad hoc and  phenomenological  way,  in previous work along this line.\cite{sud} This  led to  phenomenological constraints  on the  parameters  characterizing  the suggested collapse mechanism,  in order to ensure the theory  is able to reproduce the observational results.  In this  manuscript, we  consider  this idea anew, employing however the rather well developed CSL formalism for the description of wave function collapse.

The article is organized as follows. Sections \ref{inflation}-\ref{cmb} present the non-collapse setting, and the remaining sections invoke the collapse.

In Section \ref{inflation}, we discuss the inflationary paradigm, and give estimates of 
important numerical quantities. Section \ref{fluct} obtains the hamiltonian for the quantum perturbations on the inflaton field and presents the dynamical solutions without collapse.  Section \ref{newt} discusses the the metric perturbation on the Robertson Walker metric known as the Newtonian Potential.  The connection between classical gravity and quantum variables is obtained by invoking \textit{semi-classical gravity}, whereby the Newtonian Potential is related to  a  certain   quantum expectation value, and this important assumption is discussed. Section \ref{cmb} reviews the relation of experimentally observed quantities to the Newtonian Potential, and thereby to  those quantum expectation values.  Section \ref{CSL} then reviews the CSL formalism.   Section \ref{Application} 
adds the CSL modification to the hamiltonian evolution of the inflaton perturbations. 
 Sections \ref{P} and  \ref{X} show how it is possible to obtain agreement with the observations. Section \ref{Phy} uses this result to obtain expressions for some physical quantities, and their probabilities of realization.   Sections \ref{NewLook} and \ref{Discussion} discuss our conclusions.  

Regarding notation, we will use signature $(-+++)$ for the metric and Wald's convention for the Riemann tensor.

\section{Inflation}\label{inflation}

The starting point for the discussion of inflation is the action of a scalar field coupled to gravity.

\beq\label{actioncol}
S[\phi] = \int d^4x \sqrt{-g} \bigg[ \frac{1}{16 \pi G} R[g] -\half \nabla_a \phi \nabla_b \phi g^{ab} - V[\phi] \bigg]
\eeq

The field equations derived from the action Eq.(\ref{actioncol}) are,

\beq\label{einstein}
G^a_{b} =  8\pi G T^a_{b}
\eeq

 where $T^a_{b}$ is give by:

\beq\label{materia}
T^a_{b} =g^{ac} \partial_c \phi \partial_b \phi + \delta^a_b ( -\half g^{cd} \partial_c \phi \partial_d \phi - V(\phi) )
\eeq

Now  we  address  a  fundamental issue, how to combine quantum theory and gravitational physics.   Despite important advances, it is well known that  we  are still lacking a fully workable and  satisfactory theory of  quantum gravity.  It is also well known that a quantum gravity approach  based  on canonical quantization  leads to what is,  in effect,  a timeless theory. \cite{Time}

On the other hand, time  seems  not  only an important aspect  of any discussion of cosmology but,   also, time is needed for the CSL dynamical reduction theory which we wish to incorporate.  These   considerations 
lead  us  to  use the approach  based on semiclassical  gravity, where matter fields are treated  at the  quantum level  whereas gravitation (although quantum at the fundamental level)  is  assumed to  be  in a regime  where it can   be treated  in a classical manner to a  very good   approximation.\cite{shortcomings}  The fact that inflation  is  thought to occur at  energy   scales  well below the Planck mass lends  support to this  assumption.  (For  further   discussion on the issue  and  on the kind  of  treatment capable of    incorporating  dynamical reduction in  such a context,  see \cite{ssc}\footnote {In contrast  with  the standard  approach   where the  quantum-classical  cut is tied to  the  background-perturbation separation, 
 in the treatment developed  in  reference\cite{ssc}, the  quantum-classical cut  is  tied  to the separation of gravitation and matter  fields. In particular, the  so called  zero mode of the
  inflaton  field  is treated  at the quantum   level.}.) 
This approach differs from  that  followed  in  standard  treatments.  There (as here),  the   starting point is a
classical background for both the  gravitation and  matter fields, but  a quantum treatment  is   used  for  the perturbations of \textit{both} the metric  and the  inflaton.    In the present work we  adopt in principle the strict semiclassical approach  described above,  limiting  the quantum treatment to the perturbation  of the  inflaton field. This difference should be  kept in mind when comparing the standard  approach  to our treatment  using   the CSL  theory.  
  
  We now  proceed to  describe the  basic  setting of the problem,  starting with  characterization of  the background  metric and inflaton field, followed by treatment of the corresponding perturbed  quantities.

\subsection{The Background.}
The analysis  here will be  based, as is usual,
on separating the  metric and scalar field into a spatially homogeneous-isotropic \emph{background}  part and
 a \emph{fluctuation} part.  That is, the scalar field is written $\phi = \phi_0 + \dphi$, while the  metric is written as $g = g_0 + \delta g$, where $\phi_0$ is a function of the conformal time\footnote{Note that $a(\eta)d\eta=dt$, where $t$ is ordinary time, so if $a(\eta)\sim -1/\eta$, then $a(\eta)$ grows exponentially
  with $t$.} $\eta$ only,  and $g_0$   characterizes  the  spatially  flat Robertson Walker cosmology, i.e., 
\begin{equation}\label{RW-SP}
ds_{0}^{2} = a^2(\eta)[-d\eta ^2 + \delta_{ij}dx^i dx^j], \qquad  \phi_0 (\eta).
\end{equation}

 The   background  metric and scalar  field  are treated classically. 
 The  scalar field  and  Friedman evolution equations  are
\begin{equation}\label{backgrnd}
	\phi''_0 +2 \frac{a'}{ a}\phi'_0 +
a^2\partial_{\phi_{0}}V[\phi_{0}] =0,
 \qquad
3\bigg(\frac{ a'}{a}\bigg)^2=4\pi G  \Big[\phi'^2_0+ 2 a^2 V[\phi_0]\Big],
\end{equation} where the prime denotes derivative with respect to the conformal time $\eta$,   whose range during the inflationary era is
\textit{negative}, i.e., $\eta\in (-\infty, -\delta) $ with $\delta>0$.

The  scale factor  corresponding  to the inflationary era  of  standard inflationary cosmology which follows from Eqs.(\ref{backgrnd}) is, to a good  approximation, 
\begin{equation}\label{scalefactor}
a(\eta)= (\frac{-1}{H_I \eta} )^{1+\epsilon}.
\end{equation}
\noindent where  the Hubble constant (the expansion rate in time $t$) is 
\begin{equation}\label{Hubble}
H_I \equiv \Big [(8\pi/3) G V[\phi_0]\Big]^{1/2}.
\end{equation}

 Written in terms of $\mH \equiv a'(\eta)/a(\eta)$,  
 the so-called slow-roll parameter $\epsilon \equiv 1-\mathcal{H'}/\mathcal{H}^2$ is considered to be very small: $\epsilon \ll 1$ during 
 the inflationary stage. As for the scalar  field $\phi_0$,  the slow roll regime corresponds  to $\phi'_0= - ( a^3/3a')\partial_{\phi}V$.\footnote{This  slow roll stage,  corresponds mathematically  to  the ``terminal velocity"  of a  body subject  to a constant force in 
 addition to  a friction term.  Thus  the condition is  $\frac{\partial^2 \phi} { \partial t^2} =0$  which   corresponds  in conformal time
  to $\phi''_0 =  \mathcal{H} \phi'_0$. Using this  in the first equation 
 in (\ref{backgrnd}) leads to $\phi'_0= - ( a^3/3a')\partial_{\phi}V$ .}
Also,  it is  assumed that the  change in  the  scalar potential is so small that the Hubble parameter $H_I $ is essentially constant.  

 According to   the standard   scenario, this inflationary era is followed by a  brief reheating period in which the inflaton field (including the fluctuation field) decays, populating the universe  with
   ordinary matter fields (a process  that, for simplicity,   we  take to occur  instantaneously). This is immediately followed by the radiation-dominated era, which begins the evolution of   standard 
   hot big bang cosmology,  leading up to the present cosmological time. While the functional  form  of $a(\eta)$ changes during  these later periods, 
 that is irrelevant for our calculation, which deals solely with the inflationary era at whose end, it is assumed, the fluctuation field provides seeds of structure which are are  transformed  into anisotropies and  inhomogeneities  in the   ordinary matter distribution by the reheating conversion process.

 We  shall take inflation to start at  $\eta=-{\cal T}$,  assume that  the inflationary regime
  ends at the start of the radiation-dominated era at $\eta=-\tau$, and set  $a=1$ at the `present  cosmological time.'   

  The   effects of  the   late time physics, comprising the   physical  processes occurring between the time of reheating and the time  of decoupling (where the matter becomes transparent to the radiation),  to the extent that we shall refer  to them,
   will be  taken as fully  codified  in what are called  transfer functions (which relate the primordial fluctuations  to those  fluctuations that  are directly observable in the CMB)  and  which will be, for the most part, ignored  in the present work.

  The  only  additional information (beyond the behavior of quantities during  the inflationary regime) we  need,  in  order to  make the estimates below,    is the relationship  between  the scale  factor  at  the end of inflation,   the scale factor at the time of decoupling and the  scale factor today.
  
      Of course, the quantities directly observed  are those that were present at the  time of the decoupling (whose conformal time is  $\eta_{D}$) which lies in the matter-dominated era.  However, we shall    evaluate such quantities at the end of inflation,  at $\eta=-\tau$. We can do this because the transfer functions allow us to go backward from   $\eta=\eta_{D}$ to 
 $\eta=-\tau$.  For example,  the famous acoustic peaks, the most noteworthy feature of the CMB power spectrum, which are related to aspects of plasma physics,  
 can be effectively subtracted out using the transfer functions.  
 The results lead back (as we shall discuss later in detail) to the  scale-free Harrison-Zeldovich  spectrum at the end of inflation.   Therefore, it is the H-Z spectrum for which our calculation must account.  Alternatively said, if the transfer  functions  were constants, we would be directly observing the H-Z spectrum  today. 
 
    \subsection{Estimates.}
    
     Now,  we  need  to   estimate  the values of  the conformal time  $\eta=- \tau $ at the end of inflation  
    and the conformal time $\eta=-{\cal T}$ at which the inflation starts.  This   may be done as follows.
 
 Recall that  the temperature of radiation, regardless of era,  scales\footnote{For  a brief period  prior to the end  of reheating (which we take here  to  occur essentially instantaneously at the end of inflation), this relationship  between the temperature  and   the scale  factor  does not hold,  because  the radiation and particle  content of  the universe  are  replenished  at  that time
 as a result of the  process of  inflaton  field decay.} like  $1/a$. 
 The  radiation  temperature today corresponds to  $2.7 ^{\hbox{ o}}K =2.4\times 10^{-13} GeV$.  We  adopt  the  widespread  assumption that  the inflation scale,  i.e.,  the effective temperature at the end of inflation,  corresponds to the Grand Unification  Theory   (GUT) scale  of about $10^{15} GeV $.  Therefore, we   estimate  $a(-\tau ) = 2.4\times 10^{-28}$. 
Next, we must find the value $-\tau$ corresponding to this scale factor. 
 
  Using  the GUT inflationary  scale for  $V\sim$(GUT scale)$^{4}$,   
  and   as we  use $c=\hbar=1$(so 1GeV$\approx10^{37}$Mpc$^{-1}$ and $G= M_{Pl}^{-2}$),  we  employ   Eq.(\ref{Hubble}) to  determine   $H_I \approx 3 (M^2_{GUT}/ M_{Pl} ) =  3 (M_{GUT}/ M_{Pl} )^2\times 10^{19} GeV. = 3 \times 10^{11} GeV. $ 
   Knowing $a(-\tau)$ and $H_I$,  we  find 
 $\tau \simeq 10^{-22}$ Mpc from Eq.(\ref{scalefactor}).  (The result can be written $\tau \simeq (M_{Pl}/ M_{GUT}) 10^{-26}$ Mpc, where $M_{Pl}$ is the  Planck mass).
 
 Next, in order  to estimate the value of  the   conformal time  at the  start of inflation, $ \eta =-{\cal T}$, we assume  that  inflation lasts, say,  $70$  e-folds (usually considered a lower bound for inflation to solve the naturalness problems).  Thus  $ {\cal T}/ \tau = a(-\tau)/ a(-{\cal T})= e^{70}\approx 10^{30} $.   Combining this   with the previous  result   $\tau \simeq 10^{-22}$ Mpc. 
   gives 
${\cal T} =10^{8} $ Mpc.
   Note that  this time  becomes larger if we  require inflation to last more than  that  number of e-folds.

Also,  we  shall need  an estimate of    the  co-moving  wave number $k=|\bf k|$ for the  modes  $\bf k$
   which are relevant for the observed CMB data.  
   
   These  are obtained   by noting that  the surface of last  scattering/decouping  is  
   at $a_{LS}= 2.7K^o / 3000K^o=  1/ 1100$  and  that the co-moving  radius $R$ of the
     last scattering  sphere  is   determined by the   requirement that the photons  that  started from  a point 
     on that sphere  at  the time of decoupling   are just reaching us today. This  gives $R\approx 2/H_0 \approx 6\times  10^3$Mpc.  
     The physical radius   of that  sphere is then $R_{Phys} = a_{LS} R\approx 5.5$Mpc..

    Now, the  scales  that  are  relevant    range from those  corresponding
    to this radius to,  say, $10^{-5}$ of this value. This   corresponds to  angular scales of $2\pi\times 10^{-5}$ which is the smallest scale that can be seen 
    with current technology in the CMB. Thus
the relevant modes  are those 
  whose physical wavelength $ a_{LS} 2\pi / k$ at   last scattering  roughly  lies  between  $R_{Phys}$  and 
 $10^{-5}R_{Phys}$.  This   gives
   $ 10^{-3} \hbox{Mpc}^{-1} \lesssim k  \lesssim 10^2 \hbox {Mpc}^{-1}$. 
   
We emphasize  that  this discussion  about the  values of $k$  is   completely independent of  the precise  functional form of $a(\eta)$   between  the  end  of inflation and the decoupling time:  the only relevant data is the relationship  between  the values of the scale  factor  at  the end of inflation and the  scale factor at  the  decoupling time.
   
 It is  clear that  the  conditions $k{\cal T}>>1$ and $k\tau<<1$, which we shall use to make approximations,  hold    by  very wide margins for the relevant modes.

\section{The Fluctuation of the Inflaton Field.}\label{fluct}

Now  we  consider the fluctuation of the inflaton field, $\dphi (\x,\eta)$. 
We start with the perturbed action up to second order  in the scalar field fluctuation,  written in term of the auxiliary field $y\equiv a \dphi$,

\beq\label{111800}
\delta S^{(2)} = \frac{1}{2}\int d\eta d^3x \bigg( y'^2 - (\nabla y)^2 + \mH^2 y^2 - 2 \mH y y' \bigg).
\eeq
\noindent The  Lagrangian  density is then,

\beq\label{111803}
\delta \mathcal{L}^{(2)} =  \frac{1}{2}\bigg(y'^2 - (\nabla y)^2 + \mH^2 y^2 - 2 \mH y y'\bigg).
\eeq

The  canonical momentum $\pi$ conjugate to $y$ is $\pi \equiv \partial \delta\mathcal{L}^{(2)}/\partial y'= y' -  \mH y$. Note that 
  ${\dphi}'=(y/a)'=\pi /a $.
With  $\mH \equiv a'(\eta)/a(\eta)$ and $a(\eta) =  - 1/H_{I}\eta $,  the non-vanishing equal-time  poisson bracket and the hamiltonian are:

\begin{equation}\label {1a}
[y({\bf x}),\pi({\bf x}')]=\delta({\bf x}-{\bf x}'),
\medspace\delta\mathcal{H}^{(2)} = \frac{1}{2}\int d{\bf x}\Big[\pi^{2}({\bf x})-\frac{2}{\eta}\pi({\bf x})y({\bf x})+({\bf\nabla}y({\bf x}))^{2}\Big].
\end{equation}
\noindent where here, and in what follows, we suppress the dependence of all variables on $\eta$.

We next focus on the individual modes of the field:
\begin{equation}\label{fourier}
y({\bf x}) = \frac{1}{(2\pi)^{3/2}} \int d{{\bf k}} y({\bf k}) e^{i{\bf k}\cdot\bf x},\quad\pi({\bf x}) = \frac{1}{(2\pi)^{3/2}} \int d{\bf k} \pi({\bf k}) e^{i{\bf k}\cdot\bf x}.
\end{equation}

\noindent In terms of $y({\bf k})$,  $\pi({\bf k})$, 
which are no longer real, the non-vanishing equal-time  poisson bracket and the hamiltonian then become

\begin{equation}\label{2a}
[y({\bf k}), \pi({\bf k}')]=\delta({\bf k}-{\bf k}'),\medspace
\delta H^{(2)}  =  \frac{1}{2}\int d{\bf k}\Big [  \pi({\bf k})\pi^{*}({\bf k}) - \frac{1}{\eta}
 [ \pi^{*}({\bf k})y({\bf k}) +  \pi({\bf k}) y^{*}({\bf k}) ]+  k^2 y({\bf k})y^{*}({\bf k})\Big ]. 
\end{equation}

In making the transition from this classical description to the quantum description,  we replace a c-number  real variable $\alpha$ by a hermitian operator  $\hat{\alpha}$, 
and replace the poisson bracket by $i^{-1}\times$(commutator bracket). (We shall work in the Schr\"odinger picture, so the operators will be time-independent, and the state vector, evolving according to Schr\"odinger's equation, describes the time evolution). Thus,  $\hat{y}({\bf x})^\dagger = \hat{y}({\bf x}),$  and therefore $\hat{y}({\bf k})^\dagger = \hat{y}({-\bf k}).$
Similarly,  $\hat{\pi}({\bf x})^\dagger = \hat{\pi}({\bf x}),$  and therefore $\hat{\pi}({\bf k})^\dagger = \hat{\pi}({-\bf k}).$
Therefore,  because the classical variables describing the modes are not real, the operators $\hat{y}({\bf k})$ and $\hat{\pi}({\bf k}) $ are not
hermitian.  In order to work with hermitian operators, we turn to  a  description in terms of symmetric and antisymmetric fields.

\subsection{Symmetric and antisymmetric fields.}

Still in   the classical theory, we write each field as the sum of  symmetric and antisymmetric parts:
\begin{equation}\label {3a}
y({\bf x})= \frac{1}{2}[y({\bf x})  + y({-\bf x})] + \frac{1}{2}[y({\bf x})  - y({-\bf x})]\equiv y_{S}({\bf x})+y_{A}({\bf x}),
\end{equation}
and similarly for $\pi({\bf x})$.  Putting (\ref{3a}) into (\ref{1a}), and using $\int d{\bf x}f_{S}({\bf x})g_{A}({\bf x})=0$, Eq. (\ref{1a}) becomes
\begin{equation}\label {4}
\delta H^{(2)} =\frac{1}{2}\int d{\bf x}\Big[\pi_{S}^{2}({\bf x})-\frac{2}{\eta}\pi_{S}({\bf x})y_{S}({\bf x})+({\bf\nabla}y_{S}({\bf x}))^{2}\Big]+\frac{1}{2}\int d{\bf x}\Big[\pi_{A}^{2}({\bf x})-\frac{2}{\eta}\pi_{A}({\bf x})y_{A}({\bf x})+({\bf\nabla}y_{A}({\bf x}))^{2}\Big],
\end{equation}
\noindent i.e., $\delta H^{(2)} =\delta H^{(2)} _{S}+\delta H^{(2)} _{A}$.

The equal-time poisson brackets of the symmetric and antisymmetric fields follow from (\ref{1a}):
\begin{equation}\label {5a}
[y_{S}({\bf x}), \pi_{S}({\bf x}')]=\frac{1}{2}[\delta({\bf x}-{\bf x}')+\delta({\bf x}+{\bf x}')], \quad[y_{A}({\bf x}), \pi_{A}({\bf x}')]=\frac{1}{2}[\delta({\bf x}-{\bf x}')-\delta({\bf x}+{\bf x}')],
\end{equation}
\noindent and $[y_{S}({\bf x}), \pi_{A}({\bf x}')]=[y_{A}({\bf x}), \pi_{S}({\bf x}')]=0$.

\subsection{Modes, symmetric case.}

Now,  we  consider the individual modes of the field,
\begin{equation}\label {6}
y_{S}({\bf x})=\frac{1}{(2\pi)^{3/2}}\int d{\bf k}e^{i{\bf k}{\bf \cdot}{\bf x}} y_{S}({\bf k}),\quad  \pi_{S}({\bf x})=\frac{1}{(2\pi)^{3/2}}\int d{\bf k}e^{i{\bf k}{\bf \cdot}{\bf x}} \pi_{S}({\bf k}).                
\end{equation}
\noindent  Because $y_{S}({\bf x}), \pi_{S}({\bf x})$ are symmetric, we get $y_{S}({-\bf k})=y_{S}({\bf k})$, $\pi_{S}(-{\bf k})=\pi_{S}({\bf k})$. Because $y_{S}({\bf x}), \pi_{S}({\bf x})$ are real, we get $y_{S}^{*}({\bf k})=y_{S}({-\bf k})$, $\pi_{S}^{*}({\bf k})=\pi_{S}({-\bf k})$.  So, $y_{S}^{*}({\bf k})=y_{S}({\bf k})$, $\pi_{S}^{*}({\bf k})=\pi_{S}({\bf k})$: these classical variables are real. Therefore, the non-vanishing poisson bracket is, using  (\ref{5a}):
\begin{eqnarray}\label {7}
[y_{S}({\bf k}), \pi_{S}({\bf k}')]&=&[y_{S}({\bf k}), \pi_{S}^{*}({\bf k}')]=\frac{1}{(2\pi)^{3}}\int d{\bf x}\int d{\bf x}'e^{-i{\bf k}{\bf \cdot}{\bf x}}e^{i{\bf k}'{\bf \cdot}{\bf x}'}[y_{S}({\bf x}), \pi_{S}({\bf x}')]\nonumber\\
&=&\frac{1}{2}[\delta({\bf k}-{\bf k}')+\delta({\bf k}+{\bf k}')].
\end{eqnarray}

 In evaluating the hamiltonian, the first term is 
\begin{eqnarray}\label {8}
\frac{1}{2}\int d{\bf x}\pi_{S}^{2}({\bf x})&=&\frac{1}{2}\int d{\bf x}\pi_{S}({\bf x})\pi_{S}^{*}({\bf x})=\frac{1}{2(2\pi)^{3}}\int d{\bf x}\int d{\bf k}\int d{\bf k}'e^{i{\bf k}{\bf \cdot}{\bf x}}e^{-i{\bf k'}{\bf \cdot}{\bf x}}\pi_{S}({\bf k})\pi_{S}({\bf k}')\nonumber\\
&=&\frac{1}{2}\int d{\bf k}\pi_{S}^{2}({\bf k})=\int_{+} d{\bf k}\pi_{S}^{2}({\bf k}).
\end{eqnarray}
\noindent In the last step,   the integral over all ${\bf k}$ is converted to the integral over the upper half ${\bf k}$-plane, since the lower half ${\bf k}$-plane makes the identical contribution.  Similar steps are to be taken for the other terms in the hamiltonian, which then becomes:
\begin{equation}\label {9}
\delta H^{(2)} _{S} = \int_{+} d{\bf k}\Big[\pi_{S}^{2}({\bf k})-\frac{2}{\eta}\pi_{S}({\bf k})y_{S}({\bf k})+k^{2}y_{S}^{2}({\bf k})\Big].
\end{equation}

Finally, in limiting to the upper half ${\bf k}$-plane, we note that the $\delta({\bf k}+{\bf k}')$ term in the poisson bracket isn't used in calculating the equations of motion, so the poisson bracket is effectively just the first term in ({\ref 7}).  However,  there is a factor of $1/2$ in the poisson bracket which makes this not quite canonical.  So, we shall define new variables
\begin{equation}\label{variables}
X_{S}({\bf k})\equiv\sqrt{2} y_{S}({\bf k}),\quad P_{S}({\bf k})\equiv\sqrt{2} \pi_{S}({\bf k}).
\end{equation}

 In terms of these variables, the equal time Poisson bracket  and hamiltonian are 
\begin{equation}\label {10}
[X_{S}({\bf k}),P_{S}({\bf k}')]=\delta({\bf k}-{\bf k}'), \delta H^{(2)} _{S}=\frac{1}{2}\int_{+} d{\bf k}\Big[ P_{S}^{2}({\bf k})-\frac{2}{\eta}P_{S}({\bf k})X_{S}({\bf k})+k^{2}X_{S}^{2}({\bf k})\Big].
\end{equation}

Lastly,  we proceed to quantize,  so Eq.(\ref{10}) becomes 
\begin{equation}\label {11}
[\hat{X}_{S}({\bf k}),\hat{P}_{S}({\bf k}')]=i\delta({\bf k}-{\bf k}'), \medspace\delta H^{(2)} _{S}=\frac{1}{2}\int_{+} d{\bf k}\Big[\hat{P}_{S}^{2}({\bf k})-\frac{1}{\eta}
[\hat{P}_{S}({\bf k})\hat{X}_{S}({\bf k})+\hat{X}_{S}({\bf k})\hat{P}_{S}({\bf k})]+k^{2}\hat{X}_{S}^{2}({\bf k})\Big].
\end{equation}
 Again, we emphasize that, while the classical variables are functions of conformal time $\eta$, we are choosing to work in the 
Schr\"odinger picture.   Thus, the operators are $\eta$-independent (except for the Hamiltonian where the $\eta$-dependence is explicit): the $\eta$ dependence 
is relegated to the behavior of the state vector.

\subsection{Modes, anti-symmetric case.}

Here  we proceed in a manner  exactly parallel to that used in  the previous case.
 
We start with,
\begin{equation}\label {12}
y_{A}({\bf x})=\frac{1}{(2\pi)^{3/2}}\int d{\bf k}e^{i{\bf k}{\bf \cdot}{\bf x}}i y_{A}({\bf k}),\quad  \pi_{A}({\bf x})=\frac{1}{(2\pi)^{3/2}}\int d{\bf k}e^{i{\bf k}{\bf \cdot}{\bf x}} i \pi_{A}({\bf k}).                
\end{equation}

\noindent  However,  as $y_{A}({\bf x}), \pi_{A}({\bf x})$ are anti-symmetric, we get $y_{A}(-{\bf k})=-y_{A}({\bf k})$, $\pi_{A}(-{\bf k})=-\pi_{A}({\bf k})$. Because $y_{A}({\bf x}), \pi_{A}({\bf x})$ are real, we get $y_{A}^{*}({\bf k})=-y_{A}({-\bf k})$, $\pi_{A}^{*}({\bf k})=-\pi_{A}({-\bf k})$.  So, $y_{A}^{*}({\bf k})=y_{A}({\bf k})$, $\pi_{A}^{*}({\bf k})=\pi_{A}({\bf k})$: the $i$-factors in the definitions of $y_{A}({\bf k}), \pi_{A}({\bf k})$ were chosen to make these real so that they become hermitian operators in the transition to quantum theory. So, although $\pi({\bf k})$ is not hermitian, it has been expressed in terms of hermitian operators:
\begin{equation}\label{PSA}
\pi({\bf k})=\pi_{S}({\bf k})+i\pi_{A}({\bf k}).
\end{equation}

The non-vanishing equal time poisson bracket is, using  (\ref{5a}):
\begin{eqnarray}\label {pb2}
[y_{A}({\bf k}), \pi_{A}({\bf k}')]&=&[y_{A}({\bf k}), \pi_{A}^{*}({\bf k}')]=\frac{1}{(2\pi)^{3}}\int d{\bf x}\int d{\bf x}'e^{-i{\bf k}{\bf \cdot}{\bf x}}e^{i{\bf k}{\bf \cdot}{\bf x}'}[y_{A}({\bf x}), \pi_{A}({\bf x}')]\nonumber\\
&=&\frac{1}{2}[\delta({\bf k}-{\bf k}')-\delta({\bf k}+{\bf k}')].
\end{eqnarray}
\noindent 

The rest of the argument goes through just as for the symmetric case. The only difference is that the sign of $\delta({\bf k}+{\bf k}')$ is positive for the symmetric case, Eq.(\ref{7}), but negative for the anti-symmetric case, Eq.(\ref{pb2}).  However, when we limit to the upper half $\bf{k}$ plane, $\delta({\bf k}+{\bf k}')$ plays no role. Therefore, after quantization, we obtain the anti-symmetric version of Eq.(\ref{11}):
\begin{equation}\label {14}
[\hat{X}_{A}({\bf k}),\hat{P}_{A}({\bf k}')]=i\delta({\bf k}-{\bf k}'), \medspace\delta H^{(2)} _{A}=\frac{1}{2}\int_{+} d{\bf k}\Big[\hat{P}_{A}^{2}({\bf k})-\frac{1}{\eta}
[\hat{P}_{A}({\bf k})\hat{X}_{A}({\bf k})+\hat{X}_{S}({\bf k})\hat{P}_{A}({\bf k})]+k^{2}\hat{X}_{A}^{2}({\bf k})\Big].
\end{equation}

In this way, the field has become a simple collection of  independent modified-harmonic oscillators, with 
each mode evolving independently.  If we define 
\begin{equation}\label{discrete}
\hat{X}\equiv\sqrt{d{\bf k}}\hat{X}_{\alpha}({\bf k}), \medspace  \hat{P}\equiv\sqrt{d{\bf k}} \hat{P}_{\alpha}({\bf k}), 
\end{equation}
\noindent with indices $ \alpha=S,A, {\bf k}$ suppressed, the commutation relations and hamiltonian for a mode characterized by $k\equiv|{\bf k}|$ is

\begin{equation}\label {workingH}
[\hat{X},\hat{P}]=i, \medspace \hat{H}_{k}=\frac{1}{2}\Big[\hat{P}^{2}-\frac{1}{\eta}
[\hat{P}\hat{X}+\hat{X}\hat{P}]+k^{2}\hat{X}^{2}\Big].
\end{equation}

\subsection{Non-Collapse Theory: Expectation Values}\label{D}

We emphasize that, in our treatment, the hamiltonian  $\hat{H}_{k}$ is not  the sole cause of the dynamics.  We  have yet  to incorporate  the  CSL modification of quantum theory (Section \ref{CSL} and beyond).  Nonetheless, it is interesting, and shall prove useful, to calculate  expectation values due to the collapse-free dynamics characterized by  $H_{k}$ alone.  

This is easy to do.  For any operator  $\hat{A}$,  $\langle \psi,t|\hat{A}|\psi,t\rangle$ satisfies
\begin{equation}\label{NCEV1}
\frac{d}{d\eta}\langle \psi,\eta|\hat{A}|\psi,\eta\rangle=-i\langle \psi,\eta|[\hat{A}, \hat{H}_{k}] |\psi,\eta\rangle.  
\end{equation}
\noindent Because $\hat{H}_{k}$ is quadratic, the set of expectation values of any power of operators form a closed set which can be solved. The initial condition is that the mode wave function is in the Bunch Davies vacuum, which is just the harmonic oscillator ground state, at the initial time $\eta=-{\cal T}$
 \begin{equation}\label{NCEV2}
\langle p|\psi,-{\cal T}\rangle ={ \frac{1}{(\pi k)^{1/4}} }e^{-p^{2}/2k}, \medspace  \langle x|\psi,-{\cal T}\rangle={(\pi/ k)^{1/4}}e^{-x^{2}k/2}.
\end{equation}

Writing  $\langle\hat{A}\rangle\equiv\langle \psi,\eta|\hat{A}|\psi,\eta\rangle$, the first order equations,  the consequent equations of motion  and their solutions are
\begin{subequations}\label{NCEV3}
\begin{eqnarray}
\frac{d}{d\eta}\langle\hat{X}\rangle&=&\langle\hat{P}\rangle-\frac{\langle\hat{X}\rangle}{\eta}, \quad
\frac{d}{d\eta}\langle\hat{P}\rangle=-k^{2}\langle\hat{X}\rangle+\frac{\langle\hat{P}\rangle}{\eta}\label{NCEV3a}\\
 \frac{d^{2}}{d\eta^{2}}\langle\hat{X}\rangle&=&-\big[k^{2}-\frac{2}{\eta^{2}}\big]\langle\hat{X}\rangle, \quad
 \frac{d^{2}}{d\eta^{2}}\langle\hat{P}\rangle=-k^{2}\langle\hat{P}\rangle\label{NCEV3b}\\
 \langle\hat{X}\rangle&=&C_{1}\frac{-i}{k}e^{ik\eta}\Big[1+\frac{i}{k\eta}\Big]+ C_{2}\frac{i}{k}e^{-ik\eta}\Big[1-\frac{i}{k\eta}\Big] , \quad
 \langle\hat{P}\rangle=C_{1}e^{ik\eta}+ C_{2}e^{-ik\eta}\label{NCEV3c}.
 \end{eqnarray}
\end{subequations}
 \noindent (It is notable that $\langle\hat{P}\rangle$ has the usual harmonic oscillator solution, even though the Hamiltonian is not the usual harmonic oscillator hamiltonian.)
 From the initial conditions Eq.(\ref{NCEV2}), we see that  that $\langle\hat{X}\rangle$ and $\langle\hat{P}\rangle$ vanish initially, $C_{1}=C_{2}=0$, so for all  $\eta$, 
 \begin{equation}\label{FO}
 \langle\hat{X}\rangle= \langle\hat{P}\rangle=0.
\end{equation}

The second order equations, with $Q\equiv\langle\hat{X^{2}}\rangle$, $R\equiv\langle\hat{P^{2}}\rangle$, $S\equiv\langle[\widehat{XP}+\widehat{PX]}\rangle$,  are 
\begin{equation}\label{NCEV4}
\frac{d}{d\eta}Q=S-\frac{2Q}{\eta}, \quad\frac{d}{d\eta}R=-k^{2}S+\frac{2R}{\eta}, \quad\frac{d}{d\eta}S=2[R-k^{2}Q].
\end{equation}
\noindent Because the algebra of the commutator brackets is identical to that of the poisson brackets, and for the classical variables the product of two solutions is the solution for the product, the same is true for the solutions of Eqs.(\ref{NCEV4}). They are the product of the solutions  Eqs.(\ref{NCEV3c}):
\begin{subequations}\label{NCEV5}
\begin{eqnarray}
Q&=&-C_{1}\frac{1}{k^{2}}e^{2ik\eta}\Big[1+\frac{i}{k\eta}\Big]^{2}-C_{2}\frac{1}{k^{2}}e^{-2ik\eta}\Big[1-\frac{i}{k\eta}\Big]^{2}+
C_{3}\frac{1}{k^{2}}\Big[1+\frac{1}{(k\eta)^{2}},\Big]\label{NCEV5a}\\
R&=&C_{1}e^{2ik\eta}+C_{2}e^{-2ik\eta}+C_{3},\label{NCEV5b}\\
 S&=&-2iC_{1}\frac{1}{k}e^{2ik\eta}\Big[1+\frac{i}{k\eta}\Big]+2iC_{2}\frac{1}{k}e^{-2ik\eta}\Big[1-\frac{i}{k\eta}\Big]+
C_{3}\frac{2}{k^{2}\eta}.\label{NCEV5c}
 \end{eqnarray}
\end{subequations}
The initial conditions are
\begin{equation}\label{NCEV6}
Q(-{\cal T})=1/2k, \quad R(-{\cal T})=k/2, \quad S(-{\cal T})=0.
\end{equation}
\noindent Assuming $k{\cal T}<<1$, we obtain $C_{1}=-C_{2}=0$, $C_{3}=k/2$ and so 
\begin{equation}\label{QRS}
Q=\langle\hat{X^{2}}\rangle=\frac{1}{2k}\Big[1+\frac{1}{(k\eta)^{2}},\Big], \quad R=\langle\hat{P^{2}}\rangle=\frac{k}{2},\quad S=\langle[\widehat{XP}+\widehat{PX]}\rangle
=\frac{1}{k\eta}.
\end{equation}

It may be noted in passing, as a consequence of Eqs.(\ref{QRS}),  that an alternative choice of  canonically conjugate variables exhibits squeezing:
\[
\langle\Bigg[\sqrt{\frac{k}{2}}\hat{X}\pm\sqrt{\frac{1}{2k}}\hat{P}\Bigg]^{2}\rangle=\frac{1}{4}\Big[1\pm\frac{1}{k\eta}\Big]^{2}+\frac{1}{4},
\]
a behavior noted\cite{Referee 1,Referee 2,kiefer} as characteristic of the evolution of  the  cosmological quantum fluctuations. Here, both variables initially ($\eta=-{\cal T}$, with 
$(k{\cal T})^{-1}$ considered negligibly small) 
have the usual harmonic oscillator ground state minimum uncertainty in position and momentum but, for a range of $\eta$, the uncertainty of one of them decreases below that value, achieving a minimum at  $k\eta=-1$\footnote {The   minima   at $k\eta=+1$  would correspond to $\eta >0$  which is  not physical.}. 

 It shall be seen that the result $\langle\hat{P^{2}}\rangle=k/2$ which, together with  Eqs.(\ref{PSA}), (\ref{variables}), (\ref{discrete}),(\ref{FO}) et. seq.  implies
\begin{eqnarray}\label{mom1}
\langle\hat\pi({\bf k})\hat\pi({\bf k}')^{*}\rangle&=&\langle(\hat\pi_{S}({\bf k})+i \hat\pi_{A}({\bf k})) (\hat\pi_{S}({\bf k}')-i \hat\pi_{A}({\bf k}'))\rangle=
   \langle\hat\pi_{S}({\bf k})\hat\pi_{S}({\bf k}')\rangle   +\langle\hat\pi_{A}({\bf k})\hat\pi_{A}({\bf k}')\rangle                                    \nonumber\\
&=&2\langle\Bigg[\frac{\hat P}{\sqrt{2d{\bf k}}}\Bigg]^{2}\rangle\delta_{{\bf k}{\bf k}'}=\langle\hat P^{2}\rangle\delta({\bf k}-{\bf k}')=\frac{k}{2}\delta({\bf k}-{\bf k}'),
\end{eqnarray}
\noindent in the usual treatment is what is cited as causing agreement between the theory of inflaton perturbations and 
the effectively  observed  H-Z spectrum of temperature fluctuations. 


\section{The Fluctuation of the Metric:  Newtonian Potential.}\label{newt}

 In order to connect with observations,  we  need  the  metric perturbation $\Psi$  known  as the Newtonian potential at the end of the inflationary period. As we have mentioned, immediately following that time, the universe entered into  the brief so-called ``reheating period",  when the inflaton field is supposed to have decayed, in a not  completely understood manner, into  the matter content of the  present universe:   dark matter, baryons,  electrons, photons, etc.  As we shall see, the Newtonian potential is related to the fluctuation (inhomogeneity) in the inflaton field, so that it ends up  being  tied to the inhomogeneities of the matter created during reheating.   During reheating,  it is supposed that the Newtonian potential did not change in any appreciable manner  from its value at the end of inflation.  Thus, this primordial Newtonian potential  can be use  to determine the ensuing evolution of the matter inhomogeneities, entailing well-known physics such as baryon acoustic oscillations,  and  other processes described by the so-called transfer functions.
 
Eventually,  the universe  evolved  to what is variously called the \textit{time of last scattering} or the \textit{time of decoupling} of the photons from the plasma, or  the \textit{time of recombination} of electrons and protons,   when the atoms  formed  and  universe suddenly became transparent to the radiation which we now detect.  At that time, the universe is considered to have been in  local  thermal equilibrium.  The quantity that we presently measure is the temperature variation as a function of coordinates on the celestial sphere, $\Delta T(\theta, \varphi)/{\bar T}$ ( $\Delta T(\theta, \varphi)\equiv T(\theta, \varphi)-{\bar T}$, where ${\bar T}$ is the mean temperature over the sky). This temperature variation is  due to the inhomogeneities in the  matter density distribution at the time of last scattering. 
  
 The satellites measure the temperature by  detecting the blackbody microwave radiation at a number of frequencies.  If $\nu$ is the frequency at the peak of the spectrum, we have the relations 
\begin{equation}\label{connection}
\frac{\Delta T}{\bar T} = \frac{\delta \nu}{\nu}  \approx \frac{1}{3}\Psi.
\end{equation}
 where the last step uses the transfer functions to strip away the physics ensuing between the end of reheating and the time of decoupling, effectively 
 considering that we are directly observing $\Psi$.  The effect of $\Psi$ is, first,  a gravitational red- or blue-shift.   Second there is an effect on the rate of expansion of the universe whose combined (Sachs-Wolfe) effect gives the 1/3 factor.\footnote{ The red/blue shift contribution is $\frac{\delta \nu}{\nu}  =\frac{\delta\sqrt{g^{(e)}_{00}}}{\sqrt{g[{(o)}_{00}}}  \approx \Psi$  where $g^{(e,o)}_{00}$  represent the `` time-time" metric components  at  the events of emission and  observation  respectively. For details about the -2/3 contribution, see\cite{weinberg} p. 139, or \cite{mukbook}.} .

The Newtonian potential is related to the fluctuation of the inflaton field as follows. When the perturbation of the Robertson Walker space-time is taken into account,     
with the appropriate choice of gauge (conformal Newton gauge), and ignoring the vector and tensor part of the metric perturbations,
 Eq.(\ref{RW-SP}) is replaced by  
\begin{equation*}\label{metric}
ds^2 = a(\eta)^2 [-(1+2\Phi) d\eta^2 + (1-2\Psi) \delta_{ij} dx^idx^j],
\end{equation*}
\noindent where $\Phi$ and $\Psi$ are   functions  of  the space-time coordinates $\eta, x^i$.

Next, consider Einstein's equations to first order in the perturbations.   The expression for the energy-momentum tensor $T^a_{b}$ for the inflaton field is Eq.(\ref{materia}).  Its linear perturbation components are:
\beq\label{Teqs}
\delta T^0_0 = a^{-2}[\phi_0'^2 \Phi - \phi_0' \dphi' - \partial_\phi V a^2 \dphi], \qquad \delta T^0_i = \partial_i (-a^{-2} \phi_0' \dphi), \qquad \delta T^i_j= a^{-2}[  \phi_0' \dphi' -\phi_0'^2 \Phi - \partial_\phi V a^2 \dphi]\delta^i_j.
\eeq
\noindent Then, Einstein's equations to first order, $\delta G = 8 \pi G \delta T$, lead to $\Psi = \Phi$ and
\beq\label{nabla2psi}
\nabla^2 \Psi + \mu \Psi = 4\pi G ( \omega \dphi + \phi_0' \dphi')
\eeq

\noindent where $\mu \equiv \mH^2 - \mH'$ and $\omega \equiv 3 \mH \phi_0' + a^2 \partial_\phi V$.

 As discussed following Eq.(\ref{scalefactor}), the slow-roll approximation 
 corresponds to $ \omega=0$. We are  ignoring  here  terms of order  $\epsilon$,  which implies $\mu=0$.  Thus Eq.(\ref{nabla2psi}) and its Fourier  transform become:
\beq\label{25b}
\nabla^2 \Psi(\eta,{\bf x})  =  4 \pi G \phi_0'(\eta) \dphi'(\eta,{\bf x})= \frac{4 \pi G \phi_0'(\eta)}{a}\pi({\eta,\bf x}), \medspace -k^2 \Psi(\eta,{\bf k}) =  4 \pi G \phi_0'(\eta) \dphi'(\eta, {\bf k})= \frac{4 \pi G \phi_0'(\eta)}{a} \pi(\eta,{\bf k}).
\eeq

Now we  make the transition to quantum theory.
The usual procedure is to quantize both sides of  Eq.(\ref{25b}), so that  
$\hat\Psi({\bf k})\sim\hat\pi({\bf k})$. But then, as a consequence of Eq.(\ref{25b}) and Eq.(\ref{FO}), $\langle\hat\Psi({\bf k})\rangle\sim\langle\hat\pi({\bf k})\rangle=0$.  
Therefore, $\langle\psi,\eta|\hat\Psi({\bf x})|\psi,\eta\rangle=0$. 

How is this to be interpreted?  One would like to identify the expectation value of the Newtonian potential operator  with the value of the Newtonian potential in nature. However, for a state $|\psi,\eta\rangle$ representing nature,  the expectation value should vary with position.  Therefore, the state $|\psi,\eta\rangle$ does not represent nature.  It may then be considered as a superposition of possible states of nature, but there is no guideline how to determine the states in the superposition. 

Moreover, in the usual approach, $\langle\psi,\eta_{D}|\hat\Psi({\bf  x})\hat\Psi({\bf x}')|\psi,\eta_{D}\rangle $
 ($\eta_{D}$ represent the conformal time of decoupling) can be readily shown to be rotationally invariant, a function of $\hat{\bf x}\cdot\hat{\bf x}'$. Therefore, again, 
 $|\psi,\eta\rangle$ does not represent the state of our universe, but at best is some superposition of possible states. 
 
 So, as we have emphasized, the homogeneous and isotropic initial state and dynamics does not explain the observed inhomogeneity and anisotropy. 
 
Therefore,
as previously mentioned, following \cite{sud}\cite{shortcomings} we take a different approach, utilizing the semi-classical description of gravitation\cite{Moller}\cite{Rosenfeld}, where gravity is treated classically while other fields are treated in the standard quantum field theory (in curved space-time) fashion.  The classical gravity and the quantum fields are thus related by 
\begin{equation}\label{Semi}
G_{ab} = 8\pi G \bra \hat{T}_{ab} \ket.
\end{equation}

There is an immediate objection to semi-classical gravity.  Suppose a quantum experiment is performed with two possible macroscopic outcomes,  a large object being put in one or another place. Using the Schr\"odinger equation to describe this, including the apparatus, the resulting state vector describes a superposition of these two outcomes. 
Then,  $\bra \hat{T}_{ab} \ket$ is large in two places and so, according to semi-classical gravity, the gravitational field acts as if there were sources in two places.  Such an  experiment was actually performed\cite{Page} and, as expected,  the gravitational field (as measured by a Cavendish balance) saw only a source in one place. However, this objection no longer obtains if the Schr\"odinger equation is modified, a la CSL, to include collapse, since then the state vector rapidly ends up describing the object in one place only\footnote{For more  discussion about the  applicability of  semiclassical  gravity to the  problem at hand  see \cite{ssc}
 where the first  steps of a formalism    capable of incorporating collapse  of the wave function at the semiclassical level  was developed.}.

However, the resolution of this problem by invoking collapse brings on another problem.  As is well known, introducing CSL dynamical collapse violates the conservation of energy, so the divergence of the energy-momentum tensor does not vanish. In the current epoch this energy non-conservation is quite small, and in the present application it is also small compared to the retained terms, so it may practically be neglected. However, from a fundamental point of view, if the divergence of the  energy-momentum tensor does not vanish, and it is equated to the Einstein tensor, then of course the latter's divergence does not vanish either. 

Here  we  shall  take  the view that Einstein's  equations  are an emergent, approximate description of  the collective  behavior of the   fundamental  degrees of freedom  of quantum space-time, and  as  such those  equations  will not   hold   under all  circumstances. 
It has  been argued in \cite{ssc} 
that
  this  should  be considered  in   analogy  with the breakdown of  the  Navier-Stokes  characterization of a fluid.  This  can be expected to occur, not only   for   phenomena  at scales smaller  than the  mean   intermolecular distance   of the  fluid  constituents,   but  also when   there are important energy  fluxes between the micro and  macroscopic  degrees of freedom, such as when a part of the fluid undergoes a phase transition.   
  
   In the same  manner, the collapse  should be thought of  as   accompanied  by a back reaction in the  fundamental  quantum gravity   degrees of  freedom, which  are not fully represented in   the metric  characterization.  Then, a  more precise  description  would   include  a  compensating term   appearing in the Einstein's  equation. Another approach,  discussed  for instance in \cite{PearlePRA}, involves assigning energy-momentum to the stochastic field  driving the CSL dynamics such that the stress tensor then does have vanishing divergence. 
  We shall not  explore  this issue  any further in the present work.   
 
It follows from Eqs.(\ref{25b})and Eq.(\ref{Semi}) that
\begin{equation}\label{poisson}
 -k^2 \Psi(\eta,{\bf k}) =  4 \pi G \phi_0' (\eta) \bra \hat\dphi'( {\bf k}, \eta)\ket= \frac{4 \pi G \phi_0'(\eta)}{a}\bra \hat \pi({ \bf k},\eta)\ket 
\end{equation}
\noindent ($\bra \hat \pi({ \bf k},\eta)\ket\equiv\bra\psi,\eta | \hat \pi({ \bf k})|\psi,\eta\ket$).
When inflation starts at time $\eta=-{\cal T}$,  it is supposed that the state is described by the Bunch-Davies vacuum,  so 
 $\bra\psi,-{\cal T}| \hat \pi({ \bf k})|\psi,-{\cal T}\ket=0$, and the space-time is   homogeneous and isotropic.  While this would remain the case were there only hamiltonian dynamics,  the addition of CSL dynamics causes  
$ \bra\hat \pi({\bf k},\eta)\ket\neq 0$ 
thereafter.  The CSL hamiltonian depends upon a classical random function of time $w(t)$ (more precisely, one $w(t)$  for each momentum mode). Each set of such  $w(t)$'s 
gives rise to a different possible inhomogeneous and anisotropic universe.

\section{The Observational Quantities}\label{cmb}

The quantity that is measured is $\Delta T(\theta, \varphi)/{\bar T }$, which is a function of the coordinates on the celestial two-sphere.  This data  is expressed in terms of spherical harmonics as  
\begin{equation}\label{cmb1}
\frac{\Delta T(\theta, \varphi)}{\bar T}=\sum_{lm}\alpha_{lm}Y_{lm}(\theta, \varphi),\qquad \medspace\alpha_{lm}=\int d^{2}\Omega\frac{\Delta T (\theta, \varphi)}{\bar T}Y_{lm}^{*}(\theta, \varphi).
\end{equation}
\noindent We emphasize, as  already discussed,  that  we are  factoring out the  late  time physics,so `observations'  means what would be observed if the transfer functions were  constants.  

 The experimental results are usually expressed in terms of the  quantity 
\begin{equation}\label{alpha2}
C_{l}=\frac{1}{2l+1}\sum_{m}{|\alpha_{lm}|^{2}}.
\end{equation}
\noindent Then, the 'observation' is that the quantity 

\begin{equation}\label{alpha3}
OB_l \equiv l(l+1)(2l+1)^{-1} \sum_m |\alpha_{lm}|^2= l(l+1)C_{l}
\end{equation}
  is essentially independent of $ l$.  This `scale invariant'   (the reason for the name and the arcane dependence on $l$ shall be given subsequently), or `Harrison-Zel'dovich' spectrum, is what must be accounted for by the theory.
 
  Our approach  produces  explicit expressions for the quantities that are most  directly extracted  from the data. Using Eqs. (\ref{connection}),(\ref{poisson}), we obtain
  
  \begin{equation}\label{cmbB1}
\frac{\Delta T(\theta, \varphi)}{\bar T}= c \int  d^3 k e^{i\bf{k}\cdot\bf{x}}\frac{1}{k^{2}}\bra \hat \pi({ \bf k},\eta)\ket, \medspace\hbox{where}
\medspace c\equiv -\frac{4 \pi G \phi_0'(\eta)}{3a}.
\end{equation}  
\noindent Here,  $\bf{x}$  is   the coordinate of the  point on the   intersection of our past light cone  with what will eventually become the last scattering surface in the direction on the sky specified  by $\theta, \varphi$.  Then, according to Eq.(\ref{cmb1}), 
 \begin{equation}\label{cmbB2}
\alpha_{lm}= c \int d^{2}\Omega Y_{lm}^{*}(\theta, \varphi)  \int  d^3 k e^{i\bf{k}\cdot\bf{x}} \frac{1}{ k^2}\bra \hat \pi({ \bf k},\eta)\ket .
\end{equation}  
  \noindent Thus, $\alpha_{lm}$ depends upon $\bra \hat \pi({ \bf k},\eta)\ket$, a well-defined quantity in our treatment, which has a stochastic dependence, i.e. it depends upon a random function. (This differs from the  standard treatment, where no comparable expression can be given.)  The stochasticity occurs because the collapse theory gives an ensemble of possible universes (one for each possible random function, only one of which is actually realized) and the associated probabilities of realization. We shall see that there is not a large 
deviation from the mean, so we may consider that our universe is typical.

 Continuing, it follows from Eq.(\ref{cmbB2}) and the well-known expansion $e^{i\bf{k}\cdot\bf{x}}=4\pi\sum_{l,m}i^{l}j_{l}(kr)Y_{lm}(\theta, \varphi)
 Y_{lm}^{*}(\hat k)$ that   
     \begin{equation}\label{cmbB22}
\alpha_{lm} =i^{l}4\pi c  \int  d^3 k  j_l(kR_D) Y_{lm}^{*}(\hat k) \frac{1}{ k^2}\bra \hat \pi({ \bf k},\eta)\ket.  
\end{equation}  
Here,  $R_D$  is the co-moving   radius of the  last scattering sphere, so $\bf{x}=R_D(\sin(\theta)\sin(\varphi),\sin(\theta)\cos(\varphi),
\cos(\theta))$,  and $\hat k$ is the unit vector in the direction ${\bf k}=k\hat k$.  Therefore, 
  
     \begin{equation}\label{cmbB3}
|\alpha_{lm}|^2=
(4\pi c)^2  \int  d^3 k  d^3k' j_l(kR_D)  j_l(k'R_D)
Y_{lm}(\hat k) Y^*_{lm}(\hat k') \frac{1}{ k^{2}k'^{2}}  (\bra \hat \pi({ \bf k},\eta)\ket \bra \hat \pi({ \bf k'},\eta)\ket^*).
\end{equation}  
   We may consider that the average over the ensemble of possible universes fairly reflects the value obtained in our own universe.  As shall be seen, the form of our expression for the ensemble average at the end of the inflationary period is $\overline {(\bra \hat \pi({ \bf k},\eta)\ket \bra \hat \pi({ \bf k'},\eta)\ket^*)} = f(k) \delta( \bf{k}-\bf{k}')$. Then, 
 \begin{equation}\label{cmbB4}
\overline{|\alpha_{lm}|^2}=
	(4\pi c)^2  \int  d^3 k  j_{l}^{2}(kR_D) 
|Y_{lm}(\hat k) |^2 \frac{1}{ k^4}  f(k)  =(4\pi c)^2  \int_{0}^{\infty}   d k  j_l(k
R_D)^2  \frac{1}{ k^2}  f(k). 
\end{equation} 
     
     Now, we note that  if  $ f(k)  = \alpha k$  where $\alpha$  is a constant,  the   result  becomes independent of $R_D$. That is   what is referred to as   a  scale  invariant  spectrum. In that case,   
    
      \begin{equation}\label{cmbB5}
\overline{|\alpha_{lm}|^{2}}  =(4\pi c)^2 \alpha  \int_{0}^{\infty}   d x  j_l^{2}(x)  \frac{1}{ x}   = (4\pi c)^2 \alpha \frac{1}{2l (l+1)}.
\end{equation} 

Therefore, our  estimate for the quantity that is usually the focus  of the analysis  is

  \begin{equation}\label{CL}
C^{th}_{l}=\frac{1}{2l+1}\sum_{m=-l}^{l}\overline{|\alpha_{lm}|^{2}} =(4\pi c)^2 \alpha \frac{1}{2l (l+1)}.
\end{equation} 
Thus we have obtained the result that  $ l(l+1) C_l$ is constant,  independent of $l$,  in agreement with `observation,' as mentioned following  Eq(\ref{alpha3}). 
 However, as we have seen, that only occurs if 
\begin{equation}\label{alpha6}
\overline{\bra\hat\pi(k)\ket^{2}}\sim k.
\end{equation}

So, we now turn to add CSL dynamics to the dynamics discussed in Section \ref{D} (which is governed by the hamiltonian $\hat H_{k}$, Eq.(\ref{workingH})),  to see whether,
under the combined dynamics,   the ensemble of possible universes can satisfy (Eq.(\ref{alpha6})).  That requires $\overline{\langle\hat P\rangle^{2}}\sim k$.

 \section{CSL}\label{CSL}
 
 We shall be using just the simplest form of CSL, which describes collapse toward one or another eigenstates of an operator $\hat A$ with rate $\sim\lambda$. 

   As  we have seen, the relevant operators on which we should focus  our attention are  the $\hat \pi ({\bf k},  \eta)\sim \hat P$, and  their  expectation values $ \bra\hat \pi({\bf k},\eta)\ket\sim\langle\hat P\rangle \neq 0$. We  may  call $\hat P$ our `focus' operator  to differentiate  it  from the `collapse generating' operator $\hat A$. (Note that we could choose $\hat A=\hat P$, or make another choice for $\hat A$.) 
 
  There are two equations we must consider.   
 
 The first is a modified Schr\"odinger equation, whose  solution is:
  \begin{equation}\label{CSL1}
|\psi,t\rangle={\cal T}e^{-\int_{0}^{t}dt'\big[i\hat H+\frac{1}{4\lambda}[w(t')-2\lambda\hat A]^{2}\big]}|\psi,0\rangle.
\end{equation}
(${\cal T}$ is the time-ordering operator). $w(t)$ is a random classical function of time, of white noise type, whose probability is given by the second equation, the Probability Rule:
  \begin{equation}\label{CSL2}
	PDw(t)\equiv\langle\psi,t|\psi,t\rangle\prod_{t_{i}=0}^{t}\frac{dw(t_{i})}{\sqrt{ 2\pi\lambda/dt}}.
\end{equation}
\noindent The state vector norm evolves dynamically (does \textit{not} equal 1), so Eq.(\ref{CSL2}) says that the state vectors with largest norm are most probable.
 That the the total probability is 1 can be seen from 
 \begin{eqnarray}\label{CSL3}
\int PDw(t)&=&\int Dw(t-dt)\int_{-\infty}^{\infty}\frac{dw(t)}{\sqrt{ 2\pi\lambda/dt}}\langle\psi,t-dt| e^{-dt'\big[-i\hat H+\frac{1}{4\lambda}[w(t')-2\lambda\hat A]^{2}\big]}e^{-\int_{0}^{t}dt'\big[i\hat H+\frac{1}{4\lambda}[w(t')-2\lambda\hat A]^{2}\big]}| \psi,t-dt\rangle\nonumber\\
&=&\int Dw(t-dt)\int_{-\infty}^{\infty}\frac{dw(t)}{\sqrt{ 2\pi\lambda/dt}}\langle\psi,t-dt| e^{-\frac{1}{2\lambda}[w(t')-2\lambda\hat A]^{2}} |\psi,t-dt\rangle\nonumber\\
&=&\int Dw(t-dt)\langle\psi,t-dt|\psi,t-dt\rangle= ... =\langle\psi,0|\psi,0\rangle=1.
\end{eqnarray}

 To see how the dynamics collapses to eigenstates $|a_{n}\rangle$  of $\hat A$ (assuming  $\hat H=0$), write  $|\psi,0\rangle=\sum_{n=1}^{N}c_{n}|a_{n}\rangle$ so, 
 according to Eqs.(\ref{CSL1}),(\ref{CSL2}),
 \begin{equation}\label{CSL4}
  |\psi,t\rangle= e^{-\frac{1}{4\lambda}\int_{0}^{t}w^{2}(t')}\sum_{n=1}^{N}c_{n}|a_{n}\rangle e^{B(t)a_{n}-\lambda ta_{n}^{2}},\medspace 
 P= e^{-\frac{1}{2\lambda}\int_{0}^{t}w^{2}(t')}\sum_{n=1}^{N}|c_{n}^{2}| e^{2B(t)a_{n}-2\lambda t a_{n}^{2}},
\end{equation}
\noindent where $B(t)\equiv\int_{0}^{t}dt'w(t')$.  Writing $w(t_{i})=B(t_{i}+dt) -B(t_{i})$, so $\prod dw(t_{i})= \prod dB(t_{i})$,  we can integrate $P$ over all $B(t_{i})$ except $B(t)$, obtaining the result 
\begin{equation}\label{CSL5}
P'(B(t))dB(t)=\sum_{n=1}^{N}|c_{n}|^{2}\frac{dB(t)}{\sqrt{2\pi\lambda t}}e^{-\frac{1}{2\lambda t}[B(t)-2\lambda t a_{n}]^{2}}.
\end{equation}
\noindent According to Eq.(\ref{CSL5}), the probability is the sum of gaussians, each drifting by an amount $\sim a_{n}t$, but of width $ \sim\sqrt{\lambda t}$.  Therefore, 
after a while, they evolve into essentially separate gaussians.  Then, there are ranges of $B(t)$ which correspond to each possible outcome.   If 
$-K\sqrt{\lambda t}\leq B(t)-2\lambda ta_{n}\leq K\sqrt{\lambda t}$,  ($K>1$ is some suitably large number), the associated probability integrated over this range of  $B(t)$ 
is essentially $|c_{n}|^{2}$, and the state vector given by Eq.(\ref{CSL4}) is essentially $|\psi,t\rangle\sim |a_{n}\rangle$.

It should be emphasized that, when $\hat H\neq 0$, the hamiltonian dynamics interferes with the collapse dynamics, and various behaviors may ensue. In some cases, collapse nonetheless takes place.  In some cases, a kind of stasis or equilibrium between the two competing dynamics is reached.  In other cases, the unitary and non-unitary dynamics interfere with each other in interesting ways.  

It is useful to have an expression for the density matrix which describes the ensemble of evolutions.  This is obtained from Eq.(\ref{CSL1}):
\begin{eqnarray}\label{CSL6}
\rho(t)&=&\int PDw(t)\frac{|\psi,t\rangle\langle\psi,t|}{\langle\psi,t|\psi,t\rangle}=\int Dw(t)|\psi,t\rangle\langle\psi,t|\nonumber\\
&=&\int Dw(t){\cal T}e^{-\int_{0}^{t}dt'\big[i\hat H+\frac{1}{4\lambda}[w(t')-2\lambda\hat A]^{2}}\big]|\psi,0\rangle\langle\psi,0|
e^{-\int_{0}^{t}dt'\big[-i\hat H+\frac{1}{4\lambda}[w(t')-2\lambda\hat A]^{2}\big]}\nonumber\\
&=&{\cal T}e^{-\int_{0}^{t}dt'\big[i(\hat H_{L}-\hat H_{R}]+\frac{\lambda}{2}[\hat A_{L}-\hat A_{R}]^{2}\big]}\rho(0),
\end{eqnarray}
\noindent where the subscripts $L$ and $R$ mean that the associated operators are to be put to the left or right of $\rho(0)$, and the ${\cal T}$  reverse-time-orders operators to the right of $\rho(0)$. The evolution equation for the density matrix is therefore the simplest of Lindblad equations, 
\begin{equation}\label{CSL7}
\frac{d}{dt}\rho(t)=-i[\hat H,\rho(t)]-\frac{\lambda}{2}[\hat A,[\hat A,\rho(t)]].
\end{equation}
\noindent It follows that the ensemble expectation value of an operator$\overline{\langle\hat O\rangle}=\hbox{Tr}\hat O\rho(t)$ satisfies
\begin{equation}\label{CSL8}
\frac{d}{dt}\overline{\langle\hat O\rangle}=-i\overline{[\hat O,\hat H]}-\frac{\lambda}{2}\overline{[\hat A,[\hat A,\hat O]]}.
\end{equation}

 \section{Application of CSL}\label{Application}
 
CSL  dynamics tries to collapse state vectors toward eigenstates of $\hat A$.  It  gives an ensemble of different evolutions of the state vector, each characterized
 by a different  $w(t)$.   
 It is to be applied to the modes described by  the 
 focus operator  $\hat P$, the operator $\hat X$, and hamiltonian $\hat H_{k}$ given in Eq.(\ref{workingH}). According to  Eq.(\ref{alpha6}), using the relations which gave us Eq.(\ref{mom1}),  
\begin{eqnarray}\label{ACSL1}
\overline{ \langle\hat\pi({\bf k})\rangle\langle \hat\pi({\bf k}')\rangle^{*}}&=&
\overline{\langle(\hat\pi_{S}({\bf k})+i \hat\pi_{A}({\bf k}))\rangle\langle (\hat\pi_{S}({\bf k}')-i \hat\pi_{A}({\bf k}'))\rangle}=
\overline{\langle\hat\pi_{S}({\bf k})\rangle\langle\hat\pi_{S}({\bf k}')\rangle}  +\overline{\langle\hat\pi_{A}({\bf k})\rangle\langle\hat\pi_{A}({\bf k}')\rangle}   \nonumber\\
&=&2\overline{\langle\frac{\hat P}{\sqrt{2d{\bf k}}}\rangle^{2}}\delta_{{\bf k}{\bf k}'}=\overline{\langle\hat P\rangle^{2}}\delta({\bf k}-{\bf k}'),
 \end{eqnarray}
  \noindent what we need to find is the ensemble average $\overline{\langle\hat P\rangle^{2}}$, and determine under what circumstances, if any, this is $\sim k$. 
We must therefore choose a
collapse generating operator 
$\hat A$.  We shall consider the simplest possibilities in this paper, $\hat A= \hat X$ and $\hat A= \hat P$. These correspond to 
the basic inflaton perturbation operators $\delta\varphi({\bf x})$, $\delta\varphi'({\bf x})$.

Using Eq.(\ref{CSL8}), one can readily obtain a set of coupled equations for the ensemble average of the expectation value of any power of operators, just as was done in Section \ref{D}.  However, the ensemble average of a \textit{product} of expectation values, in particular:
\begin{equation}\label{ACSL2}
\overline{\langle\hat P\rangle^{2}}\equiv\int PDw(\eta)\frac{\langle\psi, \eta|\hat P|\psi,\eta\rangle^{2}}{\langle\psi, \eta|\psi,\eta\rangle^{2}}=\int Dw(\eta)\frac{\langle\psi, \eta|\hat P|\psi,\eta\rangle^{2}}{\langle\psi, \eta|\psi,\eta\rangle}
\end{equation}
\noindent can not be obtained in this way, and is not so easy to calculate directly.  However,  for this problem,  there is a relationship between $\overline{\langle\hat P\rangle^{2}}$ and 
$\overline{\langle\hat P^{2}\rangle}$ which allows us to obtain the former more easily.

Because the initial state is a gaussian and the hamiltonian and collapse hamiltonian are quadratic in $\hat X, \hat P$,  the form of the state vector in the momentum basis at any time is 
\begin{equation}\label{ACSL3}
\langle p|\psi,\eta \rangle = 
e^{-A(\eta)p^{2}+B(\eta)p +C(\eta)}. 
\end{equation}

The initial conditions are $A(-{\cal T})=1/2k$, $B(-{\cal T})=C(-{\cal T})=0$.  (By  solving the Schrodinger equation with this ansatz, which we shall
 eventually do,  one finds that $A$ is independent of $w(t)$, $B$ is linear in $w(t)$ and $C$ is quadratic in $w(t)$.) Therefore,  
the momentum matrix element is

\begin{eqnarray}\label{ACSL4}
\langle\psi,\eta|\hat P|\psi,\eta\rangle&=&\int dp p e^{-(A+A^{*})p^{2}+(B+B^{*})p +(C+C^{*})}\nonumber\\
&=&\frac{1}{2\sqrt{(A+A^{*}) 
}}\frac{(B+B^{*})}{(A+A^{*})}e^{\frac{(B+B^{*})^{2}}{4(A+A^{*})}}e^{(C+C^{*})}.
\end{eqnarray}
\noindent 
We also note that the state vector norm is 
\begin{equation}\label {ACSL5}
\langle\psi,\eta|\psi,\eta\rangle=\int dpe^{-(A+A^{*})p^{2}+(B+B^{*})p +(C+C^{*})}
=\frac{1}{\sqrt{(A+A^{*}) 
 }}e^{\frac{(B+B^{*})^{2}}{4(A+A^{*})}}e^{(C+C^{*})}.
\end{equation}
\noindent Therefore, by Eq.(\ref{ACSL2})
\begin{equation}\label {ACSL6}
\overline{\langle \hat P\rangle^{2}}= \int Dw(\eta)\frac{1}{\sqrt{(A+A^{*})
 }}\Bigg[\frac{(B+B^{*})}{2(A+A^{*})}\Bigg]^{2}e^{\frac{(B+B^{*})^{2}}{4(A+A^{*})}}e^{(C+C^{*})}.
\end{equation}

	Now,  $\overline{\langle \hat P^{2}\rangle}$ can be written as:
\begin{eqnarray}\label {ACSL7}
\overline{\langle \hat P^{2}\rangle}&=&\int Dw \langle\psi,\eta|\hat P^{2}|\psi,\eta\rangle\nonumber\\
&=&
\int Dw\int dp p^{2}e^{-(A+A^{*})p^{2}+(B+B^{*})p +(C+C^{*})}\nonumber\\
&=&
\int Dw \frac{1}{\sqrt{(A+A^{*})
}}\Bigg[\frac{1}{2(A+A^{*})}+\Big[\frac{(B+B^{*})}{2(A+A^{*})}\Big]^{2}\Bigg]
e^{\frac{(B+B^{*})^{2}}{4(A+A^{*})}}e^{(C+C^{*})}\nonumber\\
&=&\frac{1}{2(A+A^{*})}+\overline{\langle \hat P\rangle^{2}}
\end{eqnarray}
\noindent where the last step follows from $(A+A^{*})$ being independent of $w(t)$, from the integral of Eq.(\ref{ACSL5})  being 1 as it is the probability of all 
possible $w(t)$'s (Eq.(\ref{CSL3})), and from Eq.(\ref{ACSL6}).

To summarize,  
\begin{equation}\label {ACSL8}
\overline{\langle \hat P\rangle^{2}}=\overline{\langle \hat P^{2}\rangle}-\frac{1}{2(A+A^{*})}, 
\end{equation}
\noindent i.e.,  $[2(A+A^{*})]^{-1}$ is the standard deviation of the squared momentum.  It is also the width of every packet in momentum space.  

Thus, to calculate $\overline{\langle \hat P\rangle^{2}}$,  we shall find the second term  on the right hand side of  Eq.(\ref{ACSL8}) from the  Schr\"odinger equation and we shall find the 
first term by using the density matrix .

 \section{$\hat P$ as Generator of Collapse}\label{P}
 
\subsection{Use of Schr\"odinger equation.}
The Schr\"odinger equation is the time derivative of Eq.(\ref{CSL1}).  In the momentum representation, with $\hat A=\hat P$, it is
\begin{equation}\label{P1}
\frac{\partial}{\partial \eta}\langle p|\psi,\eta\rangle=\frac{-i}{2}\Big[p^{2}-\frac{i}{\eta}(p\frac{\partial}{\partial p}+\frac{\partial}{\partial p}p)-k^{2}\frac{\partial^{2}}{\partial p^{2}}\Big]\langle p|\psi,t\rangle-[\frac{1}{4\lambda}w^{2}(\eta)-w(\eta)p+\lambda p^{2}]\langle p|\psi,\eta\rangle,
\end{equation}
We note that $\lambda$ is a dimensionless number.  The equation $A$ satisfies is
\begin{equation}\label{P2}
\frac{d}{d\eta}A=[\frac{i}{2}+\lambda] -\frac{2}{\eta}A-2ik^{2}A^{2}.
\end{equation}
\noindent
This Ricatti equation is solved by writing $A\equiv \dot Z/[2ik^{2}Z]$. Putting this into Eq.(\ref{P2}) we get:
\begin{equation}\label{P3}
\eta\ddot Z=2ik^{2}[\frac{i}{2}+\lambda]\eta Z-2\dot Z.
\end{equation}
Defining $\alpha\equiv k\sqrt{1-2i\lambda}$, the two solutions are 
\begin{equation}\label{P4}
\frac{1}{\eta}\cos\alpha \eta \hbox{   and   }\frac{1}{\eta}\sin\alpha \eta.
\end{equation}
Therefore,
\begin{equation}\label{P5}
A(\eta)=\frac{1}{2ik^{2}\eta}\Bigg[\frac{-\cos\alpha \eta-\alpha \eta\sin\alpha \eta+C(\alpha \eta\cos\alpha \eta-\sin\alpha \eta)}{\cos \alpha\eta+C\sin \alpha \eta}\Bigg].
\end{equation}
\noindent The constant $C$ is  determined by the initial condition $A(-{\cal T})=1/2k$:  

\begin{equation}\label{P6}
C=\frac{(1-ik{\cal T})\cos\alpha {\cal T}+\alpha{\cal T}\sin\alpha {\cal T}}{(1-ik{\cal T})\sin \alpha {\cal T}-\alpha{\cal T}\cos \alpha {\cal T}}.
\end{equation} 

Putting $C$ into Eq.(\ref{P5}) yields

\begin{equation}\label{P7}
A(\eta)=\frac{i}{2k^{2}\eta}+\frac{\alpha}{2ik^{2}}\Bigg[\frac{(1-ik{\cal T})\cos\alpha( \eta+{\cal T})+\alpha{\cal T}\sin\alpha( \eta+{\cal T})}
{(1-ik{\cal T})\sin\alpha( \eta+{\cal T})-\alpha{\cal T}\cos\alpha( \eta+{\cal T})}\Bigg]\approx\frac{i}{2k^{2}\eta}+\frac{\alpha}{2k^{2}}.
\end{equation}
\noindent
The last step has utilized the approximations $k{\cal T}>>1$, $|\eta|<<{\cal T}$,  $\alpha\approx k-i\lambda k$ for $\lambda<<1$, yet $\lambda k  {\cal T}>>1$ (to be justified in section \ref{Estimates}) so  $\cos\alpha( \eta+{\cal T})\approx i\sin\alpha( \eta+{\cal T})$.
Therefore, we have the result, at $\eta=-\tau$, that 
\begin{equation}\label{P8}
\frac{1}{2(A+A^{*})}= \frac{k^{2}}{\alpha+\alpha^{*}}= \frac{k}{\sqrt{1-2i\lambda}+\sqrt{1+2i\lambda}}=  \frac{k}{\sqrt{2}\sqrt{ 1 +\sqrt{1+ 4\lambda^2}}}.
\end{equation}

\subsection{Calculation of $\overline{\langle P^{2}\rangle}$}\label{Calc}

For this case,  Eq.(\ref{CSL8}) becomes 
\begin{equation}\label{P21}
\frac{d}{d\eta}\overline{\langle\hat{\cal  O}\rangle}= -i\overline{[{\cal \hat O},\hat H]}-\frac{\lambda}{2}\overline{[\hat P,[\hat P,\hat{\cal O}]}.
\end{equation}
Referring to Section\ref{D}, where we considered the hamiltonian dynamics alone, the first order Eqs.(\ref{NCEV3}) are unchanged, so here too 
$\langle\hat X\rangle=\langle\hat P\rangle=0$.

The second order equations Eqs.(\ref{NCEV4}) 
  for $Q\equiv\overline{\langle \hat X^{2}\rangle}$, $R\equiv\overline{\langle\hat P^{2}\rangle}$, $S\equiv\overline{\langle \hat X\hat P+\hat P\hat X\rangle}$ are unchanged except for the first:  
\begin{equation}\label{P22}
\dot{Q}=S-\frac{2Q}{\eta}+\lambda,\medspace\dot{R}=-k^{2}S+\frac{2R}{\eta},\medspace
\dot{S}=2[R-k^{2} Q].
\end{equation}
\noindent The general solution is therefore the sum of the three solutions Eqs.(\ref{NCEV5}) to the homogeneous equations added to an inhomogeneous solution: 
\begin{equation}\label{P23}
Q=\lambda \eta/2,\medspace R=\lambda \eta k^{2}/2, \medspace S=\lambda/2. 
\end{equation}
\noindent For example, the equation which replaces Eq.(\ref{NCEV5b}) is
\begin{equation}\label{P24}
R=C_{1}e^{2ik\eta}+C_{2}e^{-2ik\eta}+C_{3}+\frac{\lambda k^{2}\eta}{2}.
\end{equation}
\noindent The constants are determined by the conditions at $t=-{\cal T}$, which are $Q=1/2k$, $R=k/2$, $S=0$, .  Assuming $k{\cal T}>>1$, it follows from the 
modified Eqs.(\ref{NCEV5}):
\begin{eqnarray}\label {P241}
\frac{k}{2}&=&C_{1}+C_{2}+C_{3}-\frac{\lambda k^{2}{\cal T}}{2}\nonumber\\
\frac{1}{2k}&=&-C_{1}\frac{1}{k^{2}}-C_{2}\frac{1}{k^{2}}+C_{3}\frac{1}{k^{2}}-\frac{\lambda {\cal T}}{2}\nonumber\\
0&=&-2iC_{1}\frac{1}{k}+2iC_{2}\frac{1}{k}+\frac{\lambda }{2}.
\end{eqnarray}
\noindent The solution is
\begin{equation}\label {P25}
C_{1}=-\frac{i\lambda}{8k}=-C_{2}, \quad C_{3}=\frac{k}{2}+\frac{\lambda k^{2}{\cal T}}{2}.
\end{equation}
\noindent Putting Eq.(\ref{P25}) into Eq.(\ref{P24}), setting $\eta=-\tau$, and using $k\tau<<1$, $k{\cal T}>>1$ we get:
\begin{equation}\label {P26}
R=\frac{\lambda k^{2}{\cal T}}{2} +   \frac{k}{2}.
\end{equation}
 When $\lambda k {\cal T} >> 1$ it is  clear  that the first is the dominant term, but  we include  the second term  to enable consideration of the case $\lambda\approx0$.

Therefore, by Eq.(\ref{ACSL8}), using the results (\ref{P8}) and (\ref{P26}), we obtain
\begin{equation}\label {P27}
\overline{\langle \hat P\rangle^{2}}= \frac{\lambda k^{2}{\cal T}}{2} +  \frac{k}{2} -    \frac{k}{\sqrt{2}\sqrt{ 1 +\sqrt{1+ 4\lambda^2}}}.
\end{equation}
\noindent 
If we set $\lambda=0$ (turn off CSL),  we have the  standard quantum mechanics result $\overline{\langle \hat P\rangle^{2}}=0$ since $\langle \hat P\rangle=0$. 

We see that agreement with the observed scale-invariant spectrum,  $\overline{\langle \hat P\rangle^{2}}\sim k$, can be achieved if we   assume the first  term is  dominant and we also set 
\begin{equation}\label {P28}
\lambda=\tilde\lambda/k.
\end{equation}
\noindent We note that this replaces the dimensionless collapse rate parameter $\lambda$ with parameter $\tilde\lambda$ of dimension time$^{-1}$.  

In that case  we obtain:
\begin{equation}\label {P27B}
\overline{\langle \hat P\rangle^{2}}=\frac{\tilde \lambda k{\cal T}}{2} +   \frac{k}{2} -   \frac{k}{\sqrt{2}\sqrt{ 1 +\sqrt{1+ 4(\tilde \lambda/k)^2}}}.
\end{equation}

It is also worth noting that, if $\lambda<<1$ but still 
 $\tilde\lambda {\cal T}>>1$, then
\[
\frac{\overline{\langle[ \hat P-\langle\hat P\rangle]\rangle^{2}}}{\overline{\langle \hat P^{2}\rangle}}\sim\frac{1}{\tilde\lambda {\cal T}}<<1,
\]
\noindent and the universes in the ensemble do not deviate much from each other. 

 \section{$\hat X$ as Generator of Collapse}\label{X}
 
This proceeds in a manner  parallel to the previous section. 

\subsection{Use of Schr\"odinger equation.}

  In the position representation, with $\hat A=\hat X$, the Schr\"odinger equation is
\begin{equation}\label{X1}
\frac{\partial}{\partial \eta}\langle x|\psi,\eta\rangle=\frac{-i}{2}\Big[-\frac{\partial^{2}}{\partial x^{2}}+\frac{i}{\eta}(x\frac{\partial}{\partial x}+\frac{\partial}{\partial x}x)+k^{2}x^{2}\Big]\langle x|\psi,t\rangle-[\frac{1}{4\lambda}w^{2}(\eta)-w(\eta)x+\lambda x^{2}]\langle x|\psi,\eta\rangle.
\end{equation}
We note that $\lambda$ has dimensions time$^{-2}$.  The wave function in the position representation is $\langle x|\psi,\eta\rangle=
\exp[-A'x^{2}+B'x +C']$, where $A'=1/4A$ ($A$ is the coefficient of $p^{2}$ in the exponent of the Fourier transform of  $\langle x|\psi,\eta\rangle)$satisfies
\begin{equation}\label{X2}
\frac{d}{d\eta}A'=[\frac{ik^{2}}{2}+\lambda] +\frac{2}{\eta}A'-2iA'^{2}.
\end{equation}
\noindent
This Ricatti equation is solved by writing $A'\equiv \dot Z/[2iZ]$. Putting this into Eq.(\ref{X2}) we get:
\begin{equation}\label{X3}
\eta\ddot Z=-\beta^{2}\eta Z+2\dot Z.
\end{equation}
Defining $\beta\equiv\sqrt{k^{2}-2i\lambda}$, the two solutions are 
\begin{equation}\label{X4}
e^{\pm i\beta\eta}[1\mp ib\eta].
\end{equation}
Using $\dot Z=\beta^{2}\eta\exp i\beta\eta$,  
\begin{equation}\label{X5}
A'(\eta)=\frac{\beta^{2}\eta}{2i}\Bigg[\frac{e^{2 i\beta\eta}+C}{e^{2 i\beta\eta}(1-i\beta\eta)+C(1+i\beta\eta)}\Bigg].
\end{equation}
\noindent The constant $C$ is  determined by the initial condition $A'(-{\cal T})=k/2$:   
\begin{equation}\label{X6}
C=-e^{-2 i\beta T}\frac{\beta^{2}{\cal T}-k\beta {\cal T}+ik}{\beta^{2}{\cal T}+k\beta {\cal T}+ik}.
\end{equation} 

Putting $C$ into Eq.(\ref{X5}) yields
\begin{equation}\label{X7}
A'(\eta)=\frac{\beta^{2}\eta}{2i}\Bigg[\frac{e^{2 i\beta({\cal T}+\eta)}[\beta^{2}{\cal T}+k\beta {\cal T}+ik]-[\beta^{2}{\cal T}-k\beta {\cal T}+ik]}
{e^{2 i\beta({\cal T}+\eta)}[\beta^{2}{\cal T}+k\beta {\cal T}+ik](1-i\beta\eta)-[\beta^{2}{\cal T}-k\beta {\cal T}+ik](1+i\beta\eta]}\Bigg]\approx\frac{\beta^{2}\eta}{2(i+\beta\eta)},
\end{equation}
\noindent where the last step has utilized $k{\cal T}>>1$, and $\exp 2 (-\hbox{Im}\beta){\cal T}>>1$.

Using  $A=1/4A'$, we find  
 \begin{equation}\label{X8..}
\frac{1}{2(A+A^{*})} =\frac{|A'|^{2}}{\mathrm{R}\mathrm{e}(A')}=  ( k/2)\frac{(1 + 4(\lambda/k^2)^{2})}{ F( \lambda/k^2)+ 2 ( \lambda/k^2)^2 F^{-1}(\lambda/k^2) - 2( \lambda/k^2)(k \eta)^{-1} }.
\end{equation}        
where $F(x) = \frac{1}{\sqrt{2}}\sqrt{1 + \sqrt{1 + 4x^2 }}$ ($F(0)=1$).

\subsection{Calculation of $\overline{\langle P^{2}\rangle}$}

For this case,  Eq.(\ref{CSL8}) becomes 
\begin{equation}\label{X8}
\frac{d}{d\eta}\overline{\langle\hat{\cal  O}\rangle}= -i\overline{[{\cal \hat O},\hat H]}-\frac{\lambda}{2}\overline{[\hat X,[\hat X,\hat{\cal O}]}.
\end{equation}
Referring to Section\ref{D}, where we considered the hamiltonian dynamics alone, the first order Eqs.(\ref{NCEV3}) are unchanged, so here too 
$\langle\hat X\rangle=\langle\hat P\rangle=0$.

The second order equations Eqs.(\ref{NCEV4}) 
  for $Q\equiv\overline{\langle \hat X^{2}\rangle}$, $R\equiv\overline{\langle\hat P^{2}\rangle}$, $S\equiv\overline{\langle \hat X\hat P+\hat P\hat X\rangle}$ are unchanged except for the second:  
\begin{equation}\label{X9}
\dot{Q}=S-\frac{2Q}{\eta},\medspace\dot{R}=-k^{2}S+\frac{2R}{\eta}+\lambda,\medspace
\dot{S}=2[R-k^{2} Q].
\end{equation}
\noindent The general solution is therefore the sum of the three solutions Eqs.(\ref{NCEV5}) to the homogeneous equations added to an inhomogeneous solution: 
\begin{equation}\label{X10}
Q=\lambda \eta/2k^{2},\medspace R=\lambda \eta/2, \medspace S=3\lambda/2k^{2}. 
\end{equation}
\noindent For example, the equation which replaces Eq.(\ref{NCEV5b}) is
\begin{equation}\label{X11}
R=C_{1}e^{2ik\eta}+C_{2}e^{-2ik\eta}+C_{3}+\frac{\lambda \eta}{2}.
\end{equation}
\noindent The constants are determined by the conditions at $t=-{\cal T}$, which are $Q=1/2k$, $R=k/2$, $S=0$, .  Assuming $k{\cal T}>>1$, it follows from the 
modified Eqs.(\ref{NCEV5}):
\begin{eqnarray}\label {X12}
\frac{k}{2}&=&C_{1}+C_{2}+C_{3}-\frac{\lambda {\cal T}}{2}\nonumber\\
\frac{1}{2k}&=&-C_{1}\frac{1}{k^{2}}-C_{2}\frac{1}{k^{2}}+C_{3}\frac{1}{k^{2}}-\frac{\lambda {\cal T}}{2k^{2}}\nonumber\\
0&=&-2iC_{1}\frac{1}{k}+2iC_{2}\frac{1}{k}+3\frac{\lambda }{2k^{2}}.
\end{eqnarray}
\noindent The solution is
\begin{equation}\label {X13}
C_{1}=-\frac{3i\lambda}{8k}=-C_{2}, \quad C_{3}=\frac{k}{2}+\frac{\lambda {\cal T}}{2}.
\end{equation}
\noindent Putting Eq.(\ref{X13}) into Eq.(\ref{X11}), setting $\eta=-\tau$, and using $k\tau<<1$, $\lambda{\cal T}>>k$ we get:
\begin{equation}\label {X14}
R=\frac{\lambda {\cal T}}{2}  +   \frac{k}{2} .
\end{equation}
Therefore, by Eq.(\ref{ACSL8}), using the results (\ref{X7}) and (\ref{X14}), we obtain
\begin{equation}\label {X15}
\overline{\langle \hat P\rangle^{2}} =\frac{\lambda {\cal T}}{2}-\frac{\sqrt{k^{4}+4\lambda^{2}}}{\sqrt{k^{2}-2i\lambda}+\sqrt{k^{2}+2i\lambda}}
 =\frac{\lambda {\cal T}}{2}   +   \frac{k}{2}  - ( k/2) \frac{(1 + 4(\lambda/k^2)^{2})}{ F( \lambda/k^2)+ 2 ( \lambda/k^2)^2 F^{-1}(\lambda/k^2) - 2( \lambda/k^2)(k \eta)^{-1} }.
\end{equation}
Once more, if we turn off  CSL,  we find  $\overline{\langle \hat P\rangle^{2}}=0$.

We see that agreement with the observed scale-invariant spectrum,  $\overline{\langle \hat P\rangle^{2}}\sim k$, can be achieved if we  assume that  the  first  term dominates, and  if we set 
\begin{equation}\label {X16}
\lambda=\tilde\lambda k.
\end{equation}

\noindent We note that this replaces the  collapse rate parameter $\lambda$ of dimension  time$^{-2}$ with the parameter $\tilde\lambda$ of dimension time$^{-1}$.
In that case  we obtain:

\begin{equation}\label {X15B}
\overline{\langle \hat P\rangle^{2}}=\frac{\tilde \lambda k{\cal T}}{2} +    
  \frac{k}{2}  [1-
 \frac{(1 + 4(\tilde {\lambda}/k)^{2})}{ F( \tilde {\lambda}/k)+ 2 ( \tilde {\lambda}/k)^2 F^{-1}(\tilde {\lambda}/k) - 2( \tilde {\lambda}/k)(k \eta)^{-1} }]. 
\end{equation}
The validity of our approximations shall be discussed in Section \ref{Estimates}.  

 \section{Physical Quantities}\label{Phy}
 
 As we have shown, our theory can describe  the Harrison-Zel'dovich scale-invariant spectrum.  Moreover, it can be used to find expressions for various physical quantities, 
 such as $\alpha_{lm}$, $\Delta T({\bf n})$, $\Psi({\bf x})$ (and their probabilities of taking on various values), which is 
 not possible with the usual approach. That is, we can calculate expressions for quantities corresponding to an \textit{individual universe}, not just for the ensemble of universes.
 This we shall now demonstrate,  using collapse generator  $\hat A=\hat P$  discussed in Section {\ref P}.

First, we need the expression for $\langle\psi,\eta|\hat P|\psi,\eta\rangle/\langle\psi,\eta|\psi,\eta\rangle\equiv\langle\hat P\rangle$.  To get this, we return to the Schr\"odinger equation Eq.(\ref{P1}) for $\langle p|\psi,\eta\rangle=
\exp[-Ap^{2}+Bp +C]$. With use of  Eq.(\ref{P7}) for the already-obtained variable $A$, we
find the equation for $B$:
\begin{equation}\label{RS1}
\frac{d}{d\eta}B=-\frac{1}{\eta}B-2ik^{2}AB+w\approx -\frac{1}{\eta}B-2ik^{2}\Big[\frac{i}{2k^{2}\eta}+\frac{\alpha}{2k^{2}}\Big]B+w =                                   
-i\alpha B+w\quad\hbox{or}\quad B(t)=\int_{-{\cal T}}^{\eta}d\eta'w(\eta')e^{-i\alpha(\eta-\eta')}. 
\end{equation}
\noindent Thus, according to Eqs.(\ref{ACSL4}),(\ref{ACSL5}), and (\ref{RS1}),
\begin{equation}\label{RS2}
\langle\hat P\rangle(\eta)=\frac{B+B^{*}}{2(A+A^{*})}=\frac{k^{2}}{R}\int_{-{\cal T}}^{\eta}d\eta'w(\eta')e^{-S(\eta-\eta')}
\cos R(\eta-\eta')
\end{equation}
\noindent where we have written 
\begin{equation}\label{RS3}
\alpha=k\sqrt{1-2i\lambda}\equiv R-iS,\medspace\hbox{so}\medspace R^{2}-S^{2}=k^{2}, \medspace RS=\lambda k^{2}, \hbox{ so }R=kF(\lambda), 
\medspace S=\frac{\lambda k}{F(\lambda)}.
\end{equation}

Now we can express physical quantities in terms of $\langle\hat P\rangle$.  
As was done in obtaining (\ref{ACSL1}), and using (\ref{RS2}) we write
\begin{equation}\label{RS31}
\langle\hat\pi({\bf k}, \eta)\rangle=\langle\hat\pi_{S}({\bf k},\eta)\rangle+i\langle\hat\pi_{A}({\bf k},\eta)\rangle=\frac{\langle\hat P_{S}\rangle+i\langle\hat P_{A}\rangle}{2\sqrt{d{\bf k}}}
=\frac{k^{2}}{2R}\int_{-{\cal T}}^{\eta}d\eta'[w_{S}({\bf k},\eta')+iw_{A}({\bf k},\eta')]e^{-S(\eta-\eta')}\cos R(\eta-\eta'),
\end{equation}
\noindent where we have introduced white noise functions $w({\bf k},\eta)$. These  are only defined in the upper half $ {\bf k}$-plane.  However, the fourier transform of $\hat\pi({\bf k})$ is $\hat\pi({\bf x})$ which is real.  This implies $w_{S}(-{\bf k})=w_{S}({\bf k})$ and $w_{A}(-{\bf k})=-w_{A}({\bf k})$, so  
Eq.(\ref{RS31}) holds for all ${\bf k}$.

To go along with the expression  (\ref{RS2}) for $\langle\hat P\rangle$, we must have  the probability of $w(\eta)$. To get the probability, we could  go back to Schr\'odinger's equation Eq.(\ref{P1}) and calculate $C$.  However, since the probability is $\langle\psi,\eta|\psi,\eta\rangle$, it is easier to use Eq.(\ref{P1}) 
 to get 
 \begin{eqnarray}\label{RS5}
\frac{d}{d\eta}\langle\psi,\eta|\psi,\eta\rangle&=&-\frac{w^{2}(\eta)}{2\lambda}\langle\psi,\eta|\psi,\eta\rangle+2w(\eta)\langle\psi,\eta|\hat P|\psi,\eta\rangle-2\lambda\langle\psi,\eta|\hat P^{2}|\psi,\eta\rangle \nonumber\\
&=&\Big[-\frac{w^{2}(\eta)}{2\lambda}+2w(\eta)\langle\hat P\rangle-2\lambda\langle\hat P^{2}\rangle\Big] \langle\psi,\eta|\psi,\eta\rangle
=\Big\{ -\frac{1}{2\lambda}[w(\eta)-2\lambda \langle\hat P\rangle]^{2}  -\frac{\lambda k}{F(\lambda)} \Big\}\langle\psi,\eta|\psi,\eta\rangle
\end{eqnarray}
\noindent  where the last step follows since Eq.\ref{ACSL8}) holds without the ensemble average.  Thus, we obtain the probability density 
\begin{equation}\label{RS6}
P(w)=\langle\psi,\tau|\psi,\tau\rangle=e^{-\frac{1}{2\lambda}\int_{-{\cal T}}^{\tau}d\eta[w(\eta)-2\lambda \langle\hat P\rangle(\eta)]^{2}},
\end{equation}
\noindent where $\langle\hat P\rangle(\eta)$ is given by Eq.(\ref{RS2})(the factor $\exp-\lambda k(\tau+{\cal T})/F(\lambda)$ has been absorbed in the normalization appropriate to $Dw$). 

To go to the continuum case, replace $\int d\eta$ by $\int d\eta d{\bf k}$ and $w(\eta)$ by $w({\bf k},\eta)$ in Eq.(\ref{RS6}).

In order to do calculations with the probability (\ref{RS6}), it is convenient to define a new random variable:
\begin{equation}\label{RS7}
v(\eta)\equiv w(\eta)-2\lambda \langle\hat P\rangle(\eta)=w(\eta)-2S\int_{-{\cal T}}^{\eta}d\eta'w(\eta')e^{-S(\eta-\eta')}
\cos R(\eta-\eta').
\end{equation}
\noindent The Jacobian determinant of the transformation from variables $w(\eta)$ to $v(\eta)$ is 1 (essentially, the matrix transformation has all diagonal elements =1,  and zeros to the right of the diagonal).  The probability density of $v(\eta)$ is very simple, 
\begin{equation}\label{RS8}
P(v)=e^{-\frac{1}{2\lambda}\int_{-{\cal T}}^{\tau}d\eta v(\eta)^{2}}, \medspace \overline{v(\eta)}=0,\medspace \overline{v(\eta)v(\eta')}
=\lambda\delta(\eta-\eta').
\end{equation}
\noindent However, since the physical quantities are expressed in terms of $w$, we need to invert Eq.(\ref{RS7}) to obtain the expression for 
$w$ in terms of $v$.  This is done in Appendix \ref{AppA}, with the result:
\begin{equation}\label{RS9}
w(\eta)=v(\eta)+\frac{2S}{k}\int_{-{\cal T}}^{\eta}d\eta'[S\sin k(\eta-\eta')+k\cos  k(\eta-\eta')]v(\eta').
\end{equation}

We shall provide a few examples of the use of this formalism.  We shall show, using it, that we obtain the result (\ref{P27}) for  $\overline{\langle\hat P\rangle^{2}}$.  We shall calculate the probability distribution of  the temperature  fluctuations (which shall prove to be a gaussian) as well as the correlation function of the temperature  fluctuations. Finally, we exhibit the expression for $\alpha_{lm}$, and the ensemble average of  $|\alpha_{lm}|^{2}$.

\subsection{Calculation of $\overline{\langle\hat P\rangle^{2}}$--Again.}\label{again}

As a consistency check, we employ a different way of calculating $\overline{\langle\hat P\rangle^{2}}$ than was done in Section \ref{P}.  Using 
Eqs.(\ref{RS2}) and (\ref{RS9}), and taking $\eta=0$, we  write
\begin{equation}\label{Calc1}
\langle\hat P\rangle=\frac{k^{2}}{R}\int_{-{\cal T}}^{0}d\eta'e^{S\eta'}\cos R(\eta')\Bigg[v(\eta')+\frac{2S}{k}\int_{-{\cal T}}^{\eta'}d\eta_{1}[S\sin k(\eta'-\eta_{1})+k\cos  k(\eta'-\eta_{1})]v(\eta_{1})\Bigg].    
\end{equation}
\noindent We immediately note that $\overline{\langle\hat P\rangle}=0$, since $\overline{v(\eta)}=0$.

The order of integration of the double integral can next be exchanged, and the integral over $\eta'$ performed.  There is a term which cancels the single integral in Eq.(\ref{Calc1}), and the result is 
\begin{equation}\label{Calc2}
\langle\hat P\rangle=\frac{k}{R}\int_{-{\cal T}}^{0}d\eta_{1}v(\eta_{1})\big[k\cos  k\eta_{1}-S\sin k\eta_{1}\big].    
\end{equation}
\noindent Therefore, with use of Eq.(\ref{RS8}),  and neglecting terms small compared to $k\cal{T}$, we obtain
\begin{equation}\label{Calc3}
\overline{\langle\hat P\rangle^{2}}=\frac{\lambda k^{2}}{R^{2}}\int_{-{\cal T}}^{0}d\eta_{1}\big[k\cos  k\eta_{1}-S\sin k\eta_{1}\big]^{2}\approx
  \frac{\lambda k^{2}\cal{T}}{2R^{2}}\big[S^{2}+k^{2}]=\frac{\lambda k^{2}{\cal T}}{2}.
\end{equation}
\noindent This is the same result as in Eq.(\ref{P27}).

\subsection{Temperature Fluctuation}
The temperature  fluctuation at the end of inflation is given by  Eq.(\ref{cmbB1}) with $\eta=-\tau$. (Since $k\tau<<1$, we shall replace $\tau$ by 0.)  With use of Eq.(\ref{RS31})), we therefore have 
\begin{eqnarray}\label{TF1}
\frac{\Delta T}{T}&=&c\int\frac{d{\bf k}}{k^{2}}e^{ikR_{D}{\bf \hat k}\cdot  \hat n}\langle\hat\pi({\bf k},0)\rangle=
c\int\frac{d{\bf k}}{k^{2}}e^{ikR_{D}{\bf \hat k}\cdot  \hat n}\frac{k^{2}}{2R}\int_{-{\cal T}}^{0}d\eta'[w_{S}({\bf k},\eta')+iw_{A}({\bf k},\eta')]e^{S\eta'}
\cos R\eta'\nonumber\\
&=&
c\int_{+}\frac{d{\bf k}}{k^{2}}\frac{k^{2}}{R}\int_{-{\cal T}}^{0}d\eta' e^{S\eta'}\cos R(\eta')[\cos(kR_{D}{\bf \hat k}\cdot  \hat n)w_{S}({\bf k},\eta')-
\sin (kR_{D}{\bf \hat k}\cdot  \hat n)w_{A}({\bf k},\eta')]
\end{eqnarray}
\noindent To conveniently calculate probabilities, we need to replace the $w$'s by $v$'s, using Eq.(\ref{RS9}).  Just as in the previous section, we may then exchange the order of the double integral, obtaining 
\begin{equation}\label{TF3}
\frac{\Delta T(\hat n)}{T}=
c\int_{+}\frac{d{\bf k}}{k^{2}}\frac{k}{R}\int_{-{\cal T}}^{0}d\eta_{1} [\cos(kR_{D}{\bf \hat k}\cdot  \hat n)v_{S}({\bf k},\eta_{1})-
\sin( kR_{D}{\bf \hat k}\cdot  \hat n)v_{A}({\bf k},\eta_{1})]\big[k\cos  (k\eta_{1})-S\sin (k\eta_{1})\big].
\end{equation}

\subsubsection{Probability Distribution of the Temperature Fluctuations.}

The probability that $\Delta T(\hat n)/T=C$ is given by
\begin{equation}\label{TF4}
P(C)=\int Dve^{-\int_{{\cal T}}^{0}d\eta\int_{+}d{\bf k}\frac{1}{2\lambda(k)}[v_{S}^{2}({\bf k})+v_{A}^{2}({\bf k})]}\delta(\Delta T(\hat n)/T-C).
\end{equation}
\noindent Writing Eq.(\ref{TF3}) as
\begin{equation}\label{TF5}
\frac{ \Delta T(\hat n)}{T}=\int_{-{\cal T}}^{0}d\eta_{1}\int_{+}d{\bf k}[f_{S}({\bf k},\eta_{1})v_{S}({\bf k},\eta_{1})+f_{A}({\bf k},\eta_{1})v_{A}({\bf k},\eta_{1})],
\end{equation}
\noindent Eq.(\ref{TF4}) becomes
\begin{eqnarray}\label{TF6}
P(C)&=&\int Dve^{-\int_{{\cal T}}^{0}d\eta\int_{+}d{\bf k}\frac{1}{2\lambda(k)}[v_{S}^{2}({\bf k})+v_{A}^{2}({\bf k})]}
\frac{1}{2\pi}\int_{-\infty}^{\infty}d\omega e^{i\omega(\int_{-{\cal T}}^{0}d\eta_{1}\int_{+}d{\bf k}[f_{S}({\bf k},\eta_{1})v_{S}({\bf k},\eta_{1})+f_{A}({\bf k},\eta_{1})v_{A}({\bf k},\eta_{1})]-C)}\nonumber\\
&=&\frac{1}{2\pi}\int_{-\infty}^{\infty}d\omega e^{-i\omega C}e^{-\omega^{2}\int_{-{\cal T}}^{0}d\eta_{1}\int_{+}d{\bf k}\frac{\lambda(k)}{2}
[f_{S}^{2}({\bf k},\eta_{1})+f_{A}^{2}({\bf k},\eta_{1})]}=e^{-\frac{1}{2}\frac{C^{2}}{\int_{-{\cal T}}^{0}d\eta_{1}\int_{+}d{\bf k}\lambda(k)
[f_{S}^{2}({\bf k},\eta_{1})+f_{A}^{2}({\bf k},\eta_{1})]}}.
\end{eqnarray}

Thus, we see that the probability is gaussian, with variance 
\begin{equation}\label{TF7}
\overline{\Bigg[\frac{\Delta T(\hat n)}{T}\Bigg]^{2}}=c^{2}\int_{+}d{\bf k}\lambda(k)
\frac{1}{(Rk)^{2}}\int_{-{\cal T}}^{0}d\eta_{1}\big[k\cos  (k\eta_{1})-S\sin( k\eta_{1})\big]^{2}\approx
c^{2}\int_{+}d{\bf k}\lambda(k)
\frac{{\cal T}}{2k^{2}}=
\pi c^{2}\tilde\lambda{\cal T}\int_{0}^{\infty}\frac{dk}{k}.
\end{equation}

We have set $\lambda(k)=\tilde\lambda/k$ in the observed range of $k$, $10^{-3}Mpc^{-1}<k<10^{2}Mpc^{-1}$, in order to obtain agreement with the Harrison-Zel'dovich spectrum.  
However, $\overline{[\Delta T(\hat n)/T]^{2}}$ is a measured quantity so it seems  reasonable to  expect that  the integral in  Eq.(\ref{TF7})  should  converge, and it doesn't.

 One  might   think the issue of the divergence of the above  integral could be  resolved  
 simply by  taking into account the fact that  the  temperature fluctuations   seen in the CMB,  are 
  the result of modifications by  ``late time"\footnote{That  refers  to  the regime well after inflation has  ended.} processes of the primordial seeds of cosmic structure we have calculated. The fact  however, is  that such  modifications\footnote{These  are  due to   well  understood  physics  connected  with the  behavior of  the plasmas  of ordinary  matter (quark -gluon plasma  before hadronization forms the nucleons,  nucleon-electron- photon plasma  before  nucleosynthesis, and  Hydrogen-Helium plasma  after nucleosynthesis)  codified in the so-called   transfer functions  which  describe how each mechanism alters the temperature structure. For example,  transfer functions are known to exhibit a damping effect known as  ``Silk Damping"  at short distances  (large $k$) due to  viscosity/photon diffusion.} occur  after   the  reheating  stage, and are therefore of no help at the time just prior to reheating, which  is the era to which  our  calculations    refer\footnote{Strictly speaking  at  the end of inflation  the conditions  are   still far from  thermal equilibrium, there  is therefore no    temperature and in fact no radiation. It  is only after the reheating that    the decay of the inflation  is supposed to   lead  the  system    towards a  state  of ``thermal  equilibrium". However the quantities  we are computing have  their   
  counterpart as  the  inflaton-energy-density  fluctuations  and Newtonian potential fluctuations.}. 

  If   there  were  no   modification  in   this  estimate   for the  inflationary regime    itself, 
   the resulting ensuing  fluctuations  would  be predicted to be of arbitrarily large  magnitude and a perturbative  treatment would, of course,  be invalid. 
 
The  issue  is intimately  tied  with  the  universal scale invariance of the  H-Z  spectrum and  is 
 thus  a problem that,  although often unrecognized,  afflicts  the   standard  treatment of inflationary perturbations.  
Let us  consider for instance  the  treatment  presented in  \cite{mukbook}.

   Consider the  expression for the  Newtonian potential  at a point $\bf x$.  To do  this one uses Eq.( 8.5)  of  that book  and  writes:
    \begin{equation}\label {Muck 8.5}
 \Phi({\bf x})  = \frac{1}{(2\pi)^{3/2}} \int d{\bf k} \Phi_{\bf k}e^{i{\bf k}\cdot{\bf x}} .
 \end{equation}
  The ensemble average of the square is then,  
      \begin{equation}\label {Muck 8.5c}
 \overline{ \Phi^{2}({\bf x)} } =  \frac{1}{{(2\pi)}^{3}}\int  d{\bf k}d{\bf k}'\overline{ \Phi_{\bf k} \Phi_{{\bf k}'}^{*}} e^{i({\bf k}-{\bf k}')\cdot{\bf x}} .
 \end{equation}
Next  one uses  the definition 8.11  of \cite{mukbook} which  indicates that we might  write the  ensemble average as,
 \begin{equation}\label {Muck 8.11}
\overline{ \Phi_{\bf k} \Phi_{{\bf k}'}^{*}} =  \delta({\bf k}-{\bf k}') 2\pi^2 \delta_{\Phi}^2 (k)/k^3
\end{equation}
where  in their notation $\delta_{\Phi}^2 (k)$ is the power spectrum (for the  gravitational potential $\Phi$) and is  taken to be a function only of $k \equiv ||{\bf k}||$.
Thus,  substituting in  \ref{Muck 8.5c}  we find: 
 \begin{equation}\label {Muck 8.11b}
 \overline{ \Phi^{2}({\bf x}) } = \frac{1}{4\pi} \int  d{\bf k} \delta_{\Phi}^2 (k)/k^3.
\end{equation}
  Now  we  take  eq  8.103 of \cite{mukbook}  which  indicates that  the  spectrum is  scale independent.  This requires  $\delta_{\Phi}^2 (k)$  to be   constant, independent of  $k$ and, in fact, given by $4(\varepsilon +p)/C_s$  where
  $\varepsilon$   and $p$ are  the  energy density  and pressure of the  background  inflaton field respectively, and where $C_{s}$ is the  effective ``speed of sound"   
  (see eq.  8.50 of \cite{mukbook})   which  for a canonical  scalar field is  just the sped of light. We thus  obtain:
   \begin{equation}\label {Muck 8.11c}
 \overline{ {\Phi({\bf x})}^2 } = \frac {2(\varepsilon +p)}{C_s}  \int _{0}^{\infty}   k^2 dk/k^3=  \frac {2(\varepsilon +p)}{C_s}   \int_0^\infty dk/k
\end{equation} 
which diverges just  as  our estimate of the  temperature fluctuation in \ref{TF7}. 

So, it seems that the  problem   we face  here is  not   intrinsic   to the CSL theory or, 
more generally,   to the  hypothesis  that some  kind  of   collapse   of the wave function  plays an important role  in  inflationary cosmology. What  the 
analysis in the present paper  does is show 
 the problem more explicitly.   After all what the collapse does,  roughly  speaking, is  provide a mechanism to  convert  quantum mechanical  uncertainties already there in the initial state into a range of  actual  values at a later time.
   
 Of course, if  the problem  lies  beyond the  collapse  approach and seems  intrinsic to the inflationary proposal for generation of  primordial cosmological   inhomogeneities  and  anisotropies, that   does not  mean  it can be ignored. One approach might be to consider a  similar  problem   that appears in any   treatment of  quantum fields:  the   uncertainty in the value of a field in its vacuum  state   at any point in Minkowski space-time  is  infinite. This is  often taken to indicate  that one cannot  consider such  a quantity as the  field  at a point and must,  rather, focus  attention  on   things  like  smeared  fields over space time regions.  After all,  quantum fields are  distribution-valued  operators rather  than ordinary operators  on Hilbert space. In the inflationary context however  it is  rather  unclear what  would be  the appropriate smearing one  should  consider.  
    
For our problem, we might  rather consider 
the possibility  that (\ref{TF7}) be made finite as the result of a     more complicated  behavior of $\lambda(k)\neq\tilde\lambda/k$ as $k\rightarrow 0$ and
$k\rightarrow\infty$. 
For example, we might set 
\begin{equation}\label{TF79}
\lambda(k)=\frac{\tilde\lambda}{k_{1}+k+(k^{2}/k_{2})}\medspace\hbox{so that}\medspace\int_{0}^{\infty}dk\lambda(k)\approx\tilde\lambda \ln\frac{k_{2}}{k_{1}}
\end{equation}
\noindent for $k_{2}>>k_{1}$ and the interval $(k_{2}, k_{1})$ very much wider than the range of $k$ appropriate to the CMB.  That would in effect  make a prediction, that the H-Z scale invariant spectrum not hold outside a particular range of $k$.

 On the other  hand  this  issue  might  be resolved   by  things  unrelated  to the CSL  collapse rate and  therefore  using the above   argument to modify  it  might be premature.  One possibility  that   would  certainly help in this regard  is the known  fact   that,   during the slow roll inflation, the expansion rate  is  not exactly  de-Sitter like.  This  leads, in the  usual analysis,  to a  modification of the H-Z  spectrum  corresponding to  a factor $k^{(n_s-1)}$
  where  the ``spectral index"  $n_s$ is known  both  theoretically and  empirically to   be     smaller than 1\cite{spectral Index}. 
  This would   remove the    divergence  coming  from the  $k\rightarrow\infty$  behavior.   Regarding the  divergence  arising from the 
 $k\rightarrow 0$  behavior  it seems that the  most natural resolution  would come  from noting that  the region   that undergoes inflation   would in all likelihood  have  started  as a a very   small patch, which despite the  nearly exponential  expansion resulting from inflation will  remain finite. Thus  the range of physically relevant   values of $k$
would be   bounded  from below  by some $k_{finite}>0$.

\subsubsection{Correlation of the Temperature Fluctuation.}

The correlation function of the temperature fluctuation can be found from  Eq.(\ref{TF3})
\begin{eqnarray}\label{TF8}
\overline{\frac{\Delta T(\hat n)}{T}\frac{\Delta T(\hat n')}{T}}&=&c^{2}\int_{+}d{\bf k}\lambda(k)
\frac{1}{(Rk)^{2}}\cos[kR_{D}{\bf \hat k}\cdot ( \hat n- \hat n')]\int_{-{\cal T}}^{0}d\eta_{1}\big[k\cos (k\eta_{1})-S\sin (k\eta_{1})\big]^{2}\nonumber\\
&=&4(\pi c)^{2}{\cal T}\int_{0}^{\infty}dk\lambda(k)\sum_{lm}j_{l}^{2}(kR_{D})Y_{lm}(\hat n)Y_{lm}^{*}(\hat n')\nonumber\\
&=&\pi c^{2}{\cal T}\int_{0}^{\infty}dk\lambda(k)\sum_{l}(2l+1)j_{l}^{2}(kR_{D})P_{l}(\hat n\cdot\hat n').
\end{eqnarray}
\noindent(Note, $\sum_{l}(2l+1)j_{l}^{2}(x)=1$, so Eq.(\ref{TF8}) agrees with Eq.(\ref{TF7}) when $\hat n=\hat n'$.)

\subsection{$\alpha_{lm}$}

From the definition (\ref{cmb1}) of $\alpha_{lm}$ and Eq.(\ref{TF3}) we can exhibit the expression for $\alpha_{lm}$: 
\begin{equation}\label{TF9}
\alpha_{lm}=
c\int_{+}d{\bf k}\frac{1}{kR}\int_{-{\cal T}}^{0}d\eta_{1}\sum_{l}(2l+1)j_{l}(kR_{D})P_{l}(\hat k\cdot\hat n')
[{\cal E}(l)v_{S}({\bf k},\eta_{1})+i{\cal O}(l)v_{A}({\bf k},\eta_{1})]\big[k\cos  (k\eta_{1})-S\sin (k\eta_{1})\big],
\end{equation}
\noindent where ${\cal E}(l)$ is 1 if $l$ is even and 0 if $l$ is odd, and ${\cal O}(l)$ is 0 if $l$ is even and 1 if $l$ is odd.

However, it is easiest to use the second of Eqs.(\ref{TF8}) to find 
\begin{equation}\label{TF15}
\overline{|\alpha_{lm}|^{2}}=(2\pi c)^{2}{\cal T}\int_{0}^{\infty}dk\lambda(k)j_{l}^{2}(kR_{D})=(2\pi c)^{2}{\tilde\lambda\cal T}\int_{0}^{\infty}\frac{dk}{k}j_{l}^{2}(kR_{D})=\frac{(2\pi c)^{2}{\tilde\lambda\cal T}}{2l(l+1)}.
\end{equation}
\noindent which is to be compared with Eqs.(\ref{cmbB4}), (\ref{cmbB5}).  In Eq.(\ref{TF15}) we have set $\lambda(k)=\tilde\lambda/k$, although we earlier specified that it might behave differently as $k\rightarrow 0$ and $k\rightarrow\infty$.  This different behavior has been ignored in calculating 
the integral in (\ref{TF15}), which converges perfectly well without it.  Thus,  such possibly different behavior is now constrained to have a negligible effect on the integral in Eq.(\ref{TF15}).

This concludes our demonstration that the dynamics produces agreement with the observed Harrison-Zel'dovich spectrum.

\subsection{Estimates}\label{Estimates}

  It is  necessary to make some  order of magnitude  estimates to justify the approximations we have made. We start from the  fact that  the   temperature fluctuations in the CMB  are $\frac{\Delta T}{T} = 1/3 \Psi \sim 10^{-5}$.  
We can now  use that together  with our other   results  to  check the self-consistency   of   assumptions  made in earlier sections. 
We  note  that  we took the dominant term in Eq.({\ref{P27B}) to be

  \begin{equation}\label {X15Bc}
\overline{\langle \hat P\rangle^{2}}\approx \frac{\tilde \lambda k{\cal T}}{2} .
\end{equation}
\noindent  
 This led to  the result (\ref{TF7}), 
\begin{equation}\label{TF71}
\Big[ \frac{\Delta T}{T}\Big]^{2}=\frac{\pi}{4}\Big[\frac{4\pi G\phi'_{0}}{3a}\Big]^{2}\tilde\lambda{\cal T}\cal{I}
 \end{equation}
\noindent where ${\cal I}\equiv 1/\tilde\lambda\int dk\lambda(k)$ is at least as large as $ \int_{10^{-3}}^{10^{2}} dk/k\approx 11.5$.

Using the definition   $\mH \equiv a'(\eta)/a(\eta)$  in the second of Eqs.(\ref{backgrnd}) and  taking the derivative  with respect to  $\eta$  we find 
\begin{equation}\label{aca 1}
 6\mH\mH' = 4\pi G(  2 \phi''_0\phi'_0+ 4 a a' V[\phi_0] +  2 a^2 \partial_{\phi_{0}}V[\phi_{0}]\phi'_0).
 \end{equation}
  \noindent Substituting   here the expression for  $\phi''_0$  from the first of Eqs.(\ref{backgrnd})  gives
 \begin{equation}\label{aca 2}
 6\mH\mH' = 4\pi G(  -4\mH (\phi'_0)^2+ 4 a a' V[\phi_0] ) = 16\pi G\mH (  - (\phi'_0)^2+ a^2V[\phi_0] ).
 \end{equation}
\noindent Therefore
  \begin{equation}\label{aca 3}
 \mH'  = \frac{8\pi G}{3} (  - (\phi'_0)^2+ a^2V[\phi_0] ).
 \end{equation}
 \noindent Now,  using the definition  of the slow roll parameter $\epsilon \equiv 1-\mathcal{H'}/\mathcal{H}^2$, the expression above for $\mH'$  and the second of Eqs.(\ref{backgrnd}) again to give  $\mH^2$, we have  
   \begin{equation}\label{aca 4}
\epsilon = \frac{ 3(\phi'_0)^2 }{ (\phi'_0)^2+ a^2V[\phi_0] )} \approx\frac{ 3(\phi'_0)^2 }{ ( a^2V[\phi_0] )}
  \end {equation}
 \noindent where in the last step we  have used the  standard   inflationary  requirement that the potential  term is much larger than the kinetic   term.
 Therefore  we have: $(\phi'_0)^2  \approx \epsilon a^2 V/3$.
 
 Using this result, we can rewrite the quantity $c^{2}$ appearing in  Eq.(\ref{TF71}) as
 \begin{equation}\label{aca5}
   \Big[\frac{4\pi G\phi'_{0}}{3a}\Big]^{2} =\frac{(4\pi )^2}{27}G ^2  \epsilon  V  \approx  \epsilon  \frac{V}{M_{Pl}^4} \approx  \epsilon  
 \Bigg ( \frac{M_{GUT}}{M_{Pl}}\Bigg)^4 
   \end{equation}
\noindent  Thus,  $ (\frac{\Delta T}{T})^{2} \approx \epsilon [V/ (M_{Pl})^4]\tilde\lambda{\cal T}\cal{I}$.

However as discussed in \cite{Finetunning}, the effect of reheating  (unless some very  unusual coincidences occur) is to  multiply  
this  result by  $1/ {\epsilon^2}$. Thus  we have:
\begin{equation}\label{TF10}
\Big(\frac{\Delta T}{T}\Big)^{2} \approx \frac{1}{\epsilon} \frac{V}{ (M_{Pl})^4}\tilde\lambda{\cal T}{\cal I}.
\end{equation}

The observations  yield $ (\frac{\Delta T}{T})^{2}  \sim  10^{-10}$.   The small departure from flatness of the spectrum is  taken to indicate  $ \epsilon\approx  10^{-2}$.  With $V^{1/4}$ given by the GUT scale as $\approx 10^{15}$ GeV$\approx10^{-4} M_{Pl}$, and taking  ${\cal I}\approx 10$, we conclude that 
\begin{equation}\label{TF101}
 \tilde\lambda{\cal T}\medspace \hbox{is  of order}\medspace 10^3 >> 1.
\end{equation}
\noindent In section \ref{P}, following Eq.(\ref{P7}), we used $\lambda k{\cal T}= \tilde\lambda{\cal T}>>1$, so this justifies the approximation made there.

 Since we  had estimated ${\cal T}$ to be of order $ 10^8 \hbox{Mpc}= 10^{22} \hbox{sec.}$,    we have 
\begin{equation}\label{TF102}
\tilde\lambda \approx  10^{-5} \hbox{MpC}^{-1} \approx10^{-19} \hbox{sec}^{-1}.
\end{equation}
\noindent Curiously, this is not far removed from the value $10^{-16}$sec$^{-1}$ suggested by GRW\cite{GRW} in their theory of instantaneous collapses on position, and adopted in the CSL\cite{CSL} theory of continuous dynamical collapse on mass-density. 

Thus  for the modes of interest we have  $ \tilde\lambda /k  $  in the range of $ 10^{-2}$ to $ 10^{-7}$.
We use this  estimate to  consider the magnitude of the sub-leading term in Eq.({\ref{P27B}), 

\begin{equation}\label {Est2}
\overline{\langle \hat P\rangle^{2}}=( k/2) \Bigg(\tilde \lambda {\cal T} +
 1  -  
  \frac{\sqrt{2}}{\sqrt{ 1 +\sqrt{1+ 4(\tilde \lambda/k)^2}}}\Bigg), 
\end{equation}
\noindent  where we indeed  see that it is much smaller than the leading one. 

Therefore, we have a self-consistent  assignment of values for the parameters which satisfies the approximations made and matches the theory with the observations.

 \section{Collapse  on Field Operators}\label{NewLook}
 
 We  have presented two examples  that lead to the H-Z  spectrum.  In one case, the collapse-generating operator is $\hat P\sim\hat\pi(\bf{k})$ (Section \ref{P}), whose Fourier transform is the field operator $\hat\pi({\bf{x}})$.  In the other case it is  $\hat X\sim\hat\y(\bf{k})$ (Section \ref{X}), whose Fourier transform is  the field operator $\hat\y({\bf{x}})$.  
   In this section we shall take a look at the state vector evolution written in terms of these field operators.  We shall give the argument in detail in the first case: the second case proceeds exactly similarly.
   
   The state vector evolution given by Eqs.(\ref{CSL1}), applied to the full set of commuting collapse-generating operators we have discussed, is

  \begin{eqnarray}\label{field1}
|\psi,\eta\rangle&=&{\cal T}e^{-i\int_{-{\cal T}}^{\eta}d\eta' \hat H- \int_{-{\cal T}}^{\eta}d\eta'\int_{+} d{\bf k}\frac{1}{4\lambda (k)}[w_{S}({\bf k},\eta')-2\lambda(k)\sqrt{2} \hat \pi_{S}({\bf k})]^{2}
-\int_{-{\cal T}}^{\eta}d\eta'\int_{+} d{\bf k}\frac{1}{4\lambda (k)}[w_{A}({\bf k},\eta')-2\lambda(k)\sqrt{2} \hat \pi_{A}({\bf k})]^{2}            }|\psi,0\rangle\nonumber\\
&=&{\cal T}e^{-i\int_{-{\cal T}}^{\eta}d\eta' \hat H- \frac{1}{4\tilde\lambda}\int_{-{\cal T}}^{\eta}d\eta'\int_{+} d{\bf k}[\sqrt{k}w_{S}({\bf k},\eta')-2\tilde\lambda\sqrt{2/k} \hat \pi_{S}({\bf k})]^{2}
- \frac{1}{4\tilde\lambda}\int_{-{\cal T}}^{\eta}d\eta'\int_{+} d{\bf k}[\sqrt{k}w_{A}({\bf k},\eta')-2\tilde\lambda\sqrt{2/k} \hat \pi_{A}({\bf k})]^{2}    }|\psi,0\rangle
\end{eqnarray}
\noindent where we have written $\lambda=\tilde\lambda /k$ and, following Eq.(\ref{variables}), we have written $\hat P_{S,A}({\bf k})=\sqrt{2} \hat \pi_{S,A}({\bf k})$. We now define  
$w_{S}({\bf k},t')\equiv \sqrt{2/k} w_{R}({\bf k},t')$, $w_{A}({\bf k},t')\equiv \sqrt{2/k} w_{I}({\bf k},t')$, and $w({\bf k},t')\equiv w_{R}({\bf k},t')+iw_{I}({\bf k},t')$.  We also recall that $\hat\pi({\bf k})= \pi_{S}({\bf k})+i \pi_{A}({\bf k})$.  Therefore, Eq.(\ref{field1}) may be written as
  \begin{eqnarray}\label{field2}
|\psi,\eta\rangle&=&{\cal T}e^{-i\int_{-{\cal T}}^{\eta}d\eta' \hat H-2\frac{1}{4\tilde\lambda} \int_{-{\cal T}}^{\eta}d\eta'\int_{+} d{\bf k}[w({\bf k},\eta')-2\tilde\lambda k^{-1/2} \hat \pi({\bf k})]
[w^{*}({\bf k},\eta')-2\tilde\lambda k^{-1/2} \hat \pi^{\dagger}({\bf k})]            }|\psi,-{\cal T}\rangle\nonumber\\
&=&{\cal T}e^{-i\int_{-{\cal T}}^{\eta}d\eta' \hat H-\frac{1}{4\tilde\lambda} \int_{-{\cal T}}^{\eta}d\eta'\int d{\bf k}[w({\bf k},\eta')-2\tilde\lambda k^{-1/2} \hat \pi({\bf k})]
[w^{*}({\bf k},\eta')-2\tilde\lambda k^{-1/2} \hat \pi^{\dagger}({\bf k})]            }|\psi,-{\cal T}\rangle
\end{eqnarray}
\noindent where we have replaced 2$\times$  the integral by an integral which includes the lower half ${\bf k}$-plane by defining $w_{R}(-{\bf k},t')\equiv w_{R}({\bf k},t')$,  
$w_{I}(-{\bf k},t')\equiv-w_{I}({\bf k},t')$.

We may now convert to a real noise function and a hermitian field operator  defined as
\begin{equation}\label{field3}
w({\bf x},\eta')\equiv\frac{1}{(2\pi)^{3/2}}\int d{\bf k}e^{i{\bf k}\cdot{\bf x}}w({\bf k},\eta'),\medspace 
\tilde\pi({\bf x})\equiv\frac{1}{(2\pi)^{3/2}}\int d{\bf k}e^{i{\bf k}\cdot{\bf x}}\frac{1}{k^{1/2}}\hat\pi({\bf k})=(-\nabla^{2})^{-1/4}\hat\pi({\bf x}).
\end{equation}
Putting the inverse Fourier transforms of Eqs.(\ref{field3}) into  (\ref{field2}), we get the result we have been seeking:
 \begin{equation}\label{field4}
|\psi,\eta\rangle={\cal T}e^{-i\int_{-{\cal T}}^{\eta}d\eta' \hat H-\frac{1}{4\tilde\lambda} \int_{-{\cal T}}^{\eta}d\eta'\int d{\bf x}'[w({\bf x}',\eta')-2\tilde\lambda \tilde \pi({\bf x}')]^{2}
          }|\psi,-{\cal T}\rangle.
\end{equation}
\noindent This is just the standard CSL statevector evolution, where the collapse-generating operators (toward whose joint eigenstates collapse tends) are  $\tilde \pi({\bf x})$ for all ${\bf x}$.

Similarly, in the second case, defining 
\begin{equation}\label{field5} 
\tilde y({\bf x})\equiv\frac{1}{(2\pi)^{3/2}}\int d{\bf k}e^{i{\bf k}\cdot{\bf x}}k^{1/2}y({\bf k})=(-\nabla^{2})^{1/4}\hat y({\bf x}),
\end{equation}
\noindent the result is
 \begin{equation}\label{field6}
|\psi,t\rangle={\cal T}e^{-i\int_{-{\cal T}}^{\eta}d\eta' \hat H-\frac{1}{4\tilde\lambda} \int_{-{\cal T}}^{\eta}d\eta'\int d{\bf x}[w({\bf x},\eta')-2\tilde\lambda \tilde y({\bf x})]^{2}
          }|\psi,-{\cal T}\rangle.
\end{equation}
\noindent This is just the standard CSL statevector evolution, where the collapse-generating operators (toward whose joint eigenstates collapse tends) are  $\tilde y({\bf x})$ for all ${\bf x}$.
  
Are there fundamental reasons  determining the appearance of the operators $( -\nabla^2)^{-1/4}  \hat \pi ({\bf x})$  (or
$(-\nabla^2)^{1/4}  \hat y ({\bf x})$)  rather than the more natural $ \hat \pi ({\bf x})$ (or $\hat y ({\bf x})$)?  Perhaps a truly satisfactory answer  will have to wait  for a general  theory expressing, in  all  situations, from particle  physics  to cosmology,  the  exact form of the CSL-type of modification to the evolution of quantum states.  However we might  look at   the particular  situation at hand  and offer a two part response.

First, note that the dimension of $w({\bf x}',\eta')$ and $\tilde\lambda \tilde \pi({\bf x}')$ have to be the same to achieve the CSL form exhibited in 
(\ref{field4}).  In order that the exponent in (\ref{field4}) be dimensionless, the dimension of $w({\bf x}',\eta')$ is (time)$^{-5/2}$, so $ \tilde\pi({\bf x}')$ 
must have the dimension (time)$^{-3/2}$.  Now, in order that the hamiltonian in (\ref{1a}) have the dimension (time)$^{-1}$, $ \hat\pi({\bf x}')$ has the dimension (time)$^{-2}$.  Therefore, if the collapse-generating operator is to be of the form (operator)$\times\hat\pi({\bf x}')$, that operator must have the dimension (time)$^{1/2}$, which of course is the dimension of  $( -\nabla^2)^{-1/4}$. 

But, second, then it follows that the collapse-generating operator is \textit{effectively}  $ \hat\pi({\bf x})$.  As we have seen, if we set $\hat H=0$ in Eq.(\ref{field4}),  then at infinite time the collapse  takes the system into some eigenstate of the complete commuting set of of operators $\tilde \pi({\bf x})$. But, that state will also be 
an eigenstate of the complete commuting set of of operators $\hat \pi({\bf x})$, since if $\tilde \pi({\bf x})|\phi\rangle=u({\bf x})|\phi\rangle$, then 
$\hat \pi({\bf x})|\phi\rangle=(-\nabla^{2})^{1/4}u({\bf x})|\phi\rangle$.

Therefore, we may regard the collapse-generating operator $( -\nabla^2)^{-1/4}  \hat \pi ({\bf x})$ as the theory's way of giving us, in the most natural way, collapse toward eigenstates of  the inflaton fluctuation momentum field $ \hat \pi ({\bf x})$ consistent with the CSL state vector evolution form.   It is therefore  quite remarkable  that such collapse  leads  to a prediction that agrees  with the scale  invariant  spectrum,  in agreement with the  current  cosmological observations.

A similar case may be made for Eq.(\ref{field5}) and its \textit{effective} collapse-generating operator, the  inflaton fluctuation field $\hat y ({\bf x})$.

 \section{Discussion}\label{Discussion}

In this manuscript, we treat  the emergence of the seeds of  cosmic  structure from  the dynamics of the  fluctuation of the inflaton field.  
This is the approach employed in the usual quantum treatment of this problem but,  as   previously discussed, that approach does not really account for the  emergence  of primordial inhomogeneities and anisotropies  of the universe.  Our approach differs from the standard one in that the  evolution of the state vector from the initial Bunch-Davies  vacuum to the end of  the inflation era invokes a version of the CSL  dynamical collapse theory adapted to this setting. 
We  have taken  as  the   basic   theoretical  setting  the approximation known as  semiclassical gravity. Here, gravitation is treated    at   the classical level while the
matter fields  are treated  at the quantum level, and  their energy momentum  is taken to appear in Einstein's equations  as  the corresponding  expectation value.

In this   treatment, the expectation value  of an operator differs from that given in the standard   quantum  mechanical  treatment, as a result of the CSL theory's  modification
 of  Schr\"odinger's  equation.  To apply   the CSL theory, it is necessary to  select the collapse-generating operator, toward whose eigenstates the collapse occurs.  
We  have  considered  two cases.  One is where the  collapse-generating operator represents the Fourier mode of the inflaton field  perturbation.  The other is where the collapse-generating
operator represents the Fourier mode of the  momentum  conjugate to the inflaton field  perturbation.

As we  have  indicated, the quantity of direct  observational  interest, $|\alpha_{lm}|^2$,which characterizes the  distribution of the temperature fluctuations across the celestial sphere  in terms  of  an expansion in spherical harmonics,  refers to the time of the decoupling $\eta_{D}$. Instead, we evaluated that quantity   at the end of inflation,  at $\eta=-\tau\approx 0$. The changes over this interval,  from $-\tau$ to $\eta_{D}$, are  codified in  transfer functions 
 $T_{k}(\eta_{r},\eta_{D})$:  when taken into account,  effectively give  direct observational access to the  
  spectrum at the end of inflation.  This turns out to be the Harrison-Zel'dovich scale-invariant spectrum. That is what the theory must give.  
  
 We have seen that  agreement  between observations and theory  will result if  $\overline{\langle \hat P\rangle^{2}}(k)$  turns out to be proportional  to  $k$.  This is achieved in the two cases of collapse-generating operators we considered, by choosing the collapse rate parameter $\lambda$ of the CSL theory to have a simple dependence upon the mode's momentum magnitude $k$.  
 
 In the case  where  we take  as `collapse-generating operator'  the  operator $ \hat{P}$, it is necessary that $\lambda=\tilde \lambda /k$.  $\lambda$ in this case is dimensionless, so $\tilde \lambda$ has the dimension of rate.  
 
  In the case  where  we take  as `collapse-generating operator'   the  operator $ \hat{X}$, it is necessary that $\lambda=\tilde \lambda k$.  $\lambda$ in this case has dimensions (time)$^{-2}$  so, again, 
   $\tilde \lambda$ has the dimension of rate.

If we take seriously the idea that the choice of $\lambda(k)$ at large and small values of $k$ must be chosen to make the integral in (\ref{TF7}) finite, we would expect   deviations  in the H-Z spectral form,  for large and small $k$  that,  although unobservable today,  could  be potentially  detected in  future  experiments. 

 Finally, one could ask which one of the two  options  we have discussed (or  perhaps  another one) might  be  the correct one?  Why, in order to be 
 consistent with observations (the H-Z spectrum), does the CSL parameter $\lambda$ have (in each case) a particular, simple, dependence on $k$, 
 leading to the collapse-generating operators $( -\nabla^2)^{-1/4}  \hat \pi ({\bf x})$ and  $(-\nabla^2)^{1/4}  \hat y ({\bf x})$?  
  We have no deeper reason, other than "it works out that way."  
  
One may argue, utilizing the remarks above, that  the H-Z spectrum arises from choosing $\lambda (k)$  in the simplest way consistent with the constraints that  the effective collapse parameter $\tilde \lambda$ ought to have the  dimension of rate,  that  the  evolution of each  mode  contains just one dimensional parameter, $k$, and 
 our results for $\overline {\langle\hat P\rangle^{2}}(k)$.  This is suggestive, but it is not a  fundamental or conclusive argument.    For example, 
 in the more general context of  the  inflationary problem  we  have considered,  there are other quantities 
  with the same dimension as $k$, such as the  Planck mass,  the GUT scale, the (inflationary potential)$^{1/4}$,  the mass of the inflaton field, or a combination of these.  If they are included along with $k$, there are other options to give   $\tilde \lambda$ the dimension of rate which do not yield the H-Z spectrum.   Therefore, one still wishes for a deeper reason for the choices we have  uncovered.  A response to  this  important question 
  may have to wait for the so-far phenomenologically motivated inclusion of dynamical collapse in quantum physics to be replaced by a general and fundamental theory, perhaps for a natural relationship between dynamical  collapse and  gravitation\cite{Penrose},\cite{PS3} to be uncovered.

We conclude that  the CSL theory, with a suitable choice of the operator  controlling the collapse, is capable of addressing the shortcomings  in the standard  inflationary account of the emergence of the seeds of  cosmic  structure. That is, it can choose a universe possessing inhomogeneities and anisotropies. Our results  are in agreement with observation in the regimes  so far  investigated empirically. 
We have argued that they are also consistent with and, indeed, require deviations at much smaller angular scales,  which could be uncovered  when  the  required technology   becomes available.

While this paper was being  written,  we   learned of the results  of  a  similar  study  that came to  a different conclusion\cite{J-Martin}.
We believe that the reason their  uncertainty, or spread of the final wave function for the  relevant modes ${\bf k}$,
becomes large and ours  small  is connected  with the fact that  we  are considering a different observable. 
 They looked at (or took as  `focus'  operator) the field  amplitude for the $v$ field (the Mukahnov-Sasaki  variable,  which  is a combination of  the perturbed scalar field and the   Newtonian potential).  We took as focus operator the momentum conjugate to the field  variable, as discussed at the start of  section VI. 
 
This  difference  arises due to the  different ways that   gravitation  is  considered  within the   two approaches. The work of  \cite{J-Martin} follows  the   now-traditional approach of  treating  the  gravitational perturbations just as  any other  field which  can be  subjected to  direct quantization.   The present work  follows  an  approach initiated in  \cite{sud},  which  is  based on the  idea  that here  one is dealing  with  a situation where  gravitation must be considered  as  emergent and  not  suitable  for  the  standard quantization procedure. This  view  led us to  adopt a semiclassical treatment of the gravitational perturbations (for a  more in-depth  discussion  of this point see  section VIII  of  \cite{shortcomings}). 
    
 It is  well known that  the  exponential 
  expansion   of the scale factor produces  an  extreme  squeezing in the quantum states. Expressed in terms of  suitable canonical variables, that  leads to  an enormous  growth  in  the uncertainty of the field amplitude,  and  an enormous   decrease in the  uncertainty of  the  conjugate momentum.
  The  paper  \cite{J-Martin}  focuses  on  an   amplitude operator   which, in the absence of   collapse,  exhibits  a  large  increase  in  the  uncertainty  due to the   cosmic  expansion.
On the contrary, we have been  led to  focus on the operator $\hat P$ that,  in the  absence of collapse, exhibits  no increase in the uncertainty (it remains  constant). It seems therefore that  the results can be understood  as  follows:  The   localizing effect of CSL  is not enough to overcome   the  large  increase  in  the  uncertainty of the  operator  considered  in  \cite{J-Martin} ,  but  on the  other hand  its   enough to   produce the  desired    localization in the operator whose  uncertainty  would have   remained  constant  in the  absence of  the localization effects  of CSL. This  is consistent with our finding that the fractional dispersion in $\hat P$    decreases like ${\cal T}^{-1}$.   (We  re-emphasize that the  focus operator,  the  object  for which  we compute expectation values,  should not be confused with the `collapse generating operator' $\hat A$, which is the object driving the  CSL  dynamics).
  
  In fact  we have  carried  out an  analysis  similar to  the one  performed in  sections VIII  and  IX for $\hat P$,  but  taking   the focus operator  to be
 $\hat X$. In that case  we  find  that   the   uncertainty  in the    value of    $\hat X$  for the  state that results from the CSL  evolution diverges   when the   conformal time   for the  end  of inflation $\tau \to 0$.
\begin{appendix}

\section{Finding $w(\eta)$ in terms of $v(\eta)$.}\label{AppA}

In this appendix, we  invert Eq.(\ref{RS7}),
\begin{equation}\label{A1}
v(\eta)=w(\eta)-2S\int_{-{\cal T}}^{\eta}d\eta'w(\eta')e^{-S(\eta-\eta')}
\cos R(\eta-\eta'),
\end{equation}
\noindent solving for $w(\eta)$ in terms of $v(\eta)$.  

We extend the integral's lower limit to $-\infty$, keeping in mind that  $w(\eta)=v(\eta)=0$ for $\eta<-{\cal T}$, and extend the upper limit to $\infty$ by putting a factor $\Theta(\eta-\eta')$ in the integrand ($\Theta (x)$ is the step function). Upon expressing $w(\eta')$ in the integrand in terms of its fourier transform $\tilde w(\omega)$, and changing the integration variable to $x\equiv \eta'-\eta$, we obtain
\begin{eqnarray}\label{A2}
v(\eta)&=&w(\eta)-2S\int_{-\infty}^{\infty}dx\Theta(-x)e^{Sx}\cos Rx  \int_{-\infty}^{\infty} d\omega e^{i\omega(x+\eta)} \tilde w(\omega)\nonumber\\
&=&w(\eta)-\int_{-\infty}^{\infty} d\omega e^{i\omega\eta}\frac{2S(S+i\omega)}{ (S+i\omega)^{2}+R^{2}}  \tilde w(\omega).  
\end{eqnarray}
\noindent Taking the fourier transform yields
\begin{equation}\label{A3}
\tilde v(\omega)=\tilde w(\omega)\frac{k^{2}-\omega^{2}}{ (S+i\omega)^{2}+R^{2}}\medspace\hbox{ or}\medspace \tilde w(\omega)=
\tilde v(\omega)-\frac{2S(S+i\omega)}{\omega^{2}-k^{2}}\tilde v(\omega).
\end{equation}
\noindent where we have used (\ref{RS3}). When taking the fourier transform of (\ref{A3}), the integration path is taken below the poles on the real axis, so that $w(\eta)$ depends upon $v(\eta')$ for $\eta'\leq\eta$. This results in
\begin{eqnarray}\label{A4}
w(\eta)&=&v(\eta)-2S\int_{-\infty}^{\infty}d\eta'v(\eta')\Big[S+\frac{d}{d\eta}\Big]\frac{1}{2\pi}\int_{-\infty}^{\infty}d\omega e^{i\omega(\eta-\eta')}
\frac{1}{(\omega-k-i\epsilon)(\omega+k-i\epsilon)}\nonumber\\
&=&v(\eta)+\frac{2S}{k}\int_{-\infty}^{\infty}d\eta'v(\eta')\Big[S+\frac{d}{d\eta}\Big]\Theta(\eta-\eta')\sin k(\eta-\eta'),
\end{eqnarray}
\noindent which is the result reported in Eq.(\ref{RS9}).

\section*{Acknowledgements}
  We are glad  to acknowledge  very  interesting  and useful  exchanges  with Jerome Martin,  and Vincent Vennin.
D.S.'s work  is supported in part by the
CONACYT grant No 101712.  and  by  UNAM-PAPIIT  grant IN107412. 

\end{appendix}

\end{document}